\documentclass[12pt]{iopart}
\usepackage{graphicx}
\usepackage{hyperref}
\begin{document}
\eqnobysec \title{Classical approach in quantum physics}
\author{Evgeni A. Solov'ev}

\address{Bogoliubov Laboratory of Theoretical Physics, Joint Institute for Nuclear Research, 141980 Dubna, Moscow region, Russia}

\ead{esolovev@theor.jinr.ru}
\begin{abstract}
The application of a classical approach to various quantum
problems - the secular perturbation approach to quantization of a
hydrogen atom in external fields and a helium atom, the adiabatic
switching method for calculation of a semiclassical spectrum of
hydrogen atom in crossed electric and magnetic fields, a
spontaneous decay of excited states of a hydrogen atom,
Gutzwiller's approach to Stark problem, long-lived excited states
of a helium atom recently discovered with the help of
Poincar$\acute{\mathrm{e}}$ section, inelastic transitions in slow
and fast electron-atom and ion-atom collisions -  is reviewed.
Further, a classical representation in quantum theory is
discussed. In this representation the quantum states are treating
as an ensemble of classical states. This approach opens the way to
an accurate description of the initial and final states in
classical trajectory Monte Carlo (CTMC) method and a purely
classical explanation of tunneling phenomenon. The general aspects
of the structure of the semiclassical series such as renormgroup
symmetry, criterion of accuracy and so on are reviewed as well. In
conclusion, the relation between quantum theory, classical physics
and measurement is discussed.
\end{abstract}

\tableofcontents


\section{Introduction}\label{intro}

Classical physics is the foundation-stone in understanding of a
microcosm, since all experimental devices are designed on
classical principles. A classical approach can be used as an
approximation in quantum physics also providing, sometimes, a more
adequate description of dynamics than standard quantum
approximations. The presented treatment involves all necessary
quantum properties and the word 'classical' in the title of this
paper is used just to emphasize that this review (beside Sec.5) is
based on the analysis of classical trajectories - not on the
Schr\"odinger equation in semiclassical approximation. Besides,
the classical approach to the real physical problems, for which an
accurate asymptotic description can be developed, is presented,
and original papers are cited only. The abstract problems, such as
billiard models, Feigenbaum Universality and {\it etc.}, are not
discussed here since they are rather part of mathematics than
physics.

The classical approach plays a fundamental role in quantum
mechanics. Our understanding of any result in quantum theory
relies upon classical language. For instance, since the degree of
freedom 'spin' has no analogue in classical mechanics, we have no
idea how it looks like. The classical approach has also an
advantage over different quantum approximations because of a more
adequate description of dynamics of the system.

The semiclassical quantization conditions are formulated in the
configuration space:
\begin{equation}
\int^{s_2}_{s_1}{p}(s)ds=\pi(k+\alpha_1+\alpha_2)\hbar \ , \ \ \
k=0,1,2,... \label{1.1}
\end{equation}
where $s_1$ and $s_2$ are caustics of a classical motion along the
variable $s$ of the coordinate system where the Sch\"{o}dinger
equation can be separated, and $\alpha_1$ and $\alpha_2$ are
purely quantum phase shifts (so-called Morse indices) originating
from the caustics; in the case of turning point
$\alpha=\frac{1}{4}$, in the case of Coulomb singularity
$\alpha=-\frac{1}{4}$ and in the case of the rotational motion
$\alpha=0$. In contrast to classical mechanics, the quantization
condition (\ref{1.1}) is not invariant with respect to canonical
transformations \cite{Sol10}. For instance, in action-angle
variables $\{I,\varphi\}$ obtained from the set $\{p_q,q\}$ by a
canonical transformation, the motion is always a uniform rotation
($I=const, \varphi=\omega t$) and we lose the information about
the type of caustics (turning points) which play a key role in
quantization conditions.

For the isotropic potential $V(r)$ the problem is separable in the
spherical coordinate $\{r,\vartheta,\varphi\}$ and Hamiltonian
takes the form
\begin{equation}
H=\frac{1}{2m_e}\Bigl[{p_r}^2+\frac{1}{r^2}\Bigl({p_{\vartheta}}%
^2+\frac{{p_{\varphi}}^2}{\sin^2\vartheta}\Bigr)\Bigr]+V(r),
\label{1.2}
\end{equation}
where $m_e$ is the mass of the particle, $p_r=m_e\dot r$,
$p_{\vartheta}=m_er^2\dot \vartheta$ and $p_{\varphi}=m_e
r^2\sin^2\vartheta \dot \varphi$ are canonically conjugate momenta
to coordinates $r,\ \vartheta$ and $\varphi$, respectively. Since
particle rotates uniformly along $\varphi$, $p_{\varphi}$ is
conserved, caustics are absent and quantization condition
(\ref{1.1}) gives\footnote{In (\ref{1.1}) integration corresponds
to the half-period, i.e. from $\varphi=0$ up to $\varphi=\pi$}:
\begin{equation}
p_{\varphi}=m\hbar \ , \ \ \ \ m=0,\pm1,\pm2,... , \label{1.3}
\end{equation}
The motion along $\vartheta$ variable is oscillation between two
turning points $\vartheta_1=\arcsin(m\hbar/L)$ and $\vartheta_2
=\pi-\arcsin(m\hbar/L)$, where $L$ is an angular momentum which is
obtained from the quantization condition
\begin{equation}
\int_{\vartheta_1}^{\vartheta_2}\sqrt{L^2-\frac{m^2\hbar^2}{\sin^2\vartheta}}d\vartheta=(l+1/2)\hbar
\ , \ \  l=|m|,|m|+1,... , \label{1.4}
\end{equation}
as $L=(l+1/2)\hbar$ beside one exception - $l=0$. In this case we
should accept value $L=0$ which follows from quantum
treatment\footnote{The semiclassical quantization rule is
asymptote at $l>>1$, and formally this case is beyond the validity
of semiclassical approximation.}. At last, the energy level
$E_{n_rl}$ for $l\neq 0$ is determined from the radial
quantization condition
\begin{equation}
\int_{r_1}^{r_2}\sqrt{2m_e\Bigl[E_{n_rl}-V(r)-\frac{L^2}{2m_er^2}\Bigr]}dr=(n_r+1/2)\hbar
\ , \ \  n_r=0,1,2,... . \label{1.5}
\end{equation}
The quantization condition (\ref{1.5}) gives correct value of
energies even for the low lying levels since in this case the
actual potential can be approximated by a parabolic well for which
semiclassical and quantum spectra coincide. In the case $l=0$
($L=0$), the quantization condition (\ref{1.5}) depends on the
type of inner caustic $r_1$. For Coulomb attraction
$\alpha_1=-1/4$ and it compensates the contribution from the outer
turning point $\alpha_2=1/4$ providing correct energy value.

Another general aspect is the use of the asymptotic techniques.
The semiclassical approximation or the perturbation theory is the
asymptotic series over a small parameter $\lambda$: $\sum_{n} a_n
\lambda^n$. In the first case $\lambda=\hbar^2$, and in the second
$\lambda$ is the amplitude of perturbation. Originally the
asymptotic term $\{a_n \lambda^n\}$ is the estimation of the
accuracy in the $(n-1)$-st order. At the beginning the terms $a_n
\lambda^n$ decrease with increasing $n$ up to some index
$n=n_{min}(\lambda)$. After that the terms start to increase and
we lose the accuracy. The term $a_n \lambda^n$ can be considered
as a correction only if the next term is smaller. Since the larger
$\lambda$ the smaller $n_{min}(\lambda)$ is, in application, the
leading term gives the correct result in the widest interval of
$\lambda$ and we do not need any corrections which gradually lose
the meaning of correction with increasing $\lambda$.

Atomic units ($m_e=e=\hbar=1$) are used throughout the review,
unless otherwise explicitly indicated.

\section{Energy spectrum and resonances}\label{sec:1}

\subsection{Secular perturbation theory}

In the secular perturbation theory the Hamiltonian has the form
$H({\bi p},{\bi r})=H_0({\bi p},{\bi r})+\lambda V(\bi{r})$, where
$\lambda V(\bi{r})$ is perturbation. It is assumed that the
unperturbed Hamiltonian $H_0(\bi{p},\bi{r})$ is separable. In the
$N$ dimensional case the unperturbed system has $N$ independent
integrals of motion $\Lambda_i$ ($i=1,2,...,N$), including the
energy. Under weak perturbation these integrals of motion begin to
change slowly in time. Then, the equation of motion for
$\Lambda_i(t)$ can be averaged over an unperturbed motion (at
fixed values of $\Lambda_i$) from which corrected integrals of
motion and quantization conditions are obtained.

Usually, the secular perturbation theory is formulated in
action-angle variables $\{I,\varphi\}$ which are obtained from the
set $\{p_q,q\}$ by a canonical transformation (see,
e.g.,\cite{Born34}). However, these variables are not appropriate
for quantization conditions, because after the canonical
transformation the motion is always a uniform rotation and the
information about the
type of caustics and Morse indices is lost\footnote{%
For instance, in \cite{Rich} the action-angular variables are used
to obtain the semiclassical energy spectrum in the quadratic
Zeeman effect, but semiclassical quantization condition (3.8) in
this paper has wrong Morse indices: $\alpha_1=\alpha_2=0$ instead
of $\alpha_1=\alpha_2=\frac{1}{4}$ (see equation (\ref{2.11}) in
Sec. 2.1.2) and leads to an incorrect energy spectrum (compare
table 3 in \cite{Rich} with table 1 in Sec. 2.1.2).}.

\subsubsection{Hydrogen atom in crossed electric and magnetic fields.}
When two charged particles are moving in uniform electric $\bi F$
and magnetic $\bi B$ fields the problem of separation of the
center of mass motion becomes nontrivial (see, e.g., \cite{Gor}).
The Hamiltonian of two particles with charges $Z_1=e$ and $Z_2=-e$
in external electric and magnetic fields has the form
\begin{equation}
H=\frac{1}{2m_1}\Bigl({\bi p}_1+\frac{e}{c}{\bi A}_1\Bigr)^2+
\frac{1}{2m_2}\Bigl({\bi p}_2-\frac{e}{c}{\bi A}_2\Bigr)^2 +e{\bi
F}({\bi r}_1-{\bi r}_2)-\frac{e^2}{r},
\end{equation}
where ${\bi r}={\bi r}_1-{\bi r_2}$, $c$ is the light velocity,
$m_i$, $\bi r_i$, ${\bi p}_i$, and $\bi A_i$ are the mass, the
radius-vector, the momentum and the vector-potential for $i$-th
particle ($i=1,2)$. The momentum is connected with the velocity
${\bi v}_i$ by the relation ${\bi p}_i={\bi v}_i-(e/c){\bi A}_i$.
The gauge for the vector-potential is chosen in the form ${\bi
A}_i=[{\bi B}\times{\bi r}_i]/2$, where squared brackets $[{\bi
a}\times{\bi b}]$ denote the vector multiplication of two vectors
${\bi a}$ and ${\bi b}$. In this case, instead of the total
momentum the vector ${\bi P}={\bi p}_1+{\bi p}_2-(e/2c)[{\bi
B}\times{\bi r}]$ is conserved. After separation of 'the center of
mass motion' the Hamiltonian for a relative motion of particles
takes the form
\begin{equation} \fl
H=\frac{1}{2m}{\bi p}^2-\frac{e}{2\mu c}({\bi B}\cdot[{\bi
r}\times{\bi p}])+ \frac{e^2}{8mc^2}[{\bi B}\times{\bi r}]
+e\Bigl(\bigl({\bi F}+{\bi F}_{eff}\bigr)\cdot{\bi
r}\Bigr)-\frac{e^2}{r},\label{2.2.2.2}
\end{equation}
where $M=m_1+m_2$ is the total mass, $m=m_1m_2/(m_1+m_2)$ is the
reduced mass, $\mu=m_1m_2/(m_1-m_2)$ and $({\bi a}\cdot{\bi b})$
denotes the scalar product of two vectors ${\bi a}$ and $\bi b$.
The separation constant $\bi P$ gives rise to an additional
effective electric field ${\bi F}_{eff}=[{\bi P}\times{\bi
B}]/Mc$. The appearance of the effective field ${\bi F}_{eff}$ is
the trace of gauge invariance in the two body system; it reflects
the uniform character of the homogeneous magnetic field.

The specific feature of hydrogen-like initial Hamiltonian $H_0$ is
its huge degeneracy. In classical mechanics this degeneracy is
manifested as a closed elliptic orbit of electron. In this case,
we have only one quantization condition along this orbit, and only
one quantum number is well defined. The missed quantization
conditions are determined by the perturbation. For instance, for
the  spherical perturbation, the quantization conditions are
written in the spherical coordinates, but if the perturbation is
the uniform electric field, the quantization conditions are
written in the parabolic coordinates.

The semiclassical quantization of the hydrogen atom in weak
electric and magnetic fields at arbitrary orientations of $\bi F$
and $\bi B$ was carried out by Epstein in 1923 \cite{Eps}. In this
case, the equation of motion for the electron is
\begin{equation}
\frac{d{\bi p}}{dt}=-\frac{\bi r}{r^3}-{\bi F}+\frac{1}{c}[{\bi
B}\times {\bi v}],  \label{2.1.1}
\end{equation}
where ${\bi v}$ is the velocity of the electron;  in the
nonrelativistic case it is connected with momentum as ${\bi p}=m_e
{\bi v}$. The unperturbed Kepler elliptic orbit is specified by
angular momentum $\bi L=\bi{[r\times p]}$ and Runge-Lenz vector
${\bi A}={\bi [L\times p]}+{\bi r}/r$, and can be presented in
terms of $\bi L$ and $\bi A$ as
\begin{equation} {\bi r}(t)=a[\cos \xi(t)-e]\frac{\bi
A}{A}+a\sqrt{1-e^2}\sin \xi(t)\frac{[{\bi L}\times{\bi A}]}{LA},
\end{equation}
where $a$ is the semimajor axis of the ellipse,
$e=\sqrt{1-L^2/n^2}$ is the eccentricity and $\xi$ is the Kepler
anomaly (or 'elliptic time'), which is connected with the actual
time by the relation $t=\sqrt{a}(\xi-e\sin\xi)$. Under influence
of perturbation $\bi L$ and $\bi A$ change slowly in time and the
equations of motion averaged over Kepler period read
\begin{equation}
\frac{d {\bi L}}{dt}=\frac{3}{2}n^2[{\bi F}\times{\bi
A}]+\frac{1}{2c}[{\bi B}\times{\bi L}], \ \ \  \frac{d {\bi
A}}{dt}=\frac{3}{2}[{\bi F}\times{\bi L}]+\frac{1}{2c}[{\bi
B}\times{\bi A}]. \label{2.1.3}
\end{equation}
Now let us introduce instead of $\bi L$ and $\bi A$ the new
variables ${\bi J_{1,2}}=({\bi L}\pm n{\bi A})/2$ which are
subject to the relation
\begin{equation}
{\bi J}_{1,2}^2=-\frac{1}{8H_0}=\frac{n^2}{4} \label{2.1.5}
\end{equation}
Then, equations (\ref{2.1.3}) can be rewritten in the form
\begin{equation}
\frac{d {\bi J_1}}{dt}=\frac{1}{2c}[\tilde{\bi B}_1\times{\bi
J}_1], \ \ \ \ \frac{d {\bi J_2}}{dt}=\frac{1}{2c}[\tilde{\bi
B}_2\times{\bi J}_2]. \label{2.1.4}
\end{equation}
where $\tilde{{\bi B}}_{1,2}=({\bi B}\pm 3cn{\bi F})$. Thus, the
original problem (\ref{2.1.1}) is reduced to the problem of two
independent pseudo-particles with the 'angular momenta' $\bi J_1$
and $\bi J_2$ placed in the separate effective magnetic fields
$\tilde{\bi B}_1$ and $\tilde{\bi B}_2$, i.e. the vectors $\bi
J_1$ and $\bi J_2$ uniformly rotate around the axes $\tilde{\bi
B}_{1}$ and $\tilde{\bi B}_{2}$ with frequencies
$\omega_1=|\tilde{\bi B}_{1}|/2c$ and $\omega_2=|\tilde{\bi
B}_{2}|/2c$, respectively. The quantization of subsystems like
that is quite simple. In this case, the integrals of motion are
the 'angular momenta' and projections of the 'angular momenta'
${\bi J}_{1,2}$ onto the axes $\tilde{\bi B}_{1,2}$, respectively.
In the 3D case, the quantization of angular momenta gives (see
also quantization condition (\ref{1.4}))
\begin{equation}
{\bi J}_{1,2}^2=(j+1/2)^2, \label{2.1.6}
\end{equation}
where $j$ is the 'angular' quantum number. Comparing the
right-hand sides of equations (\ref{2.1.5}) and (\ref{2.1.6}) the
following value of angular quantum number is obtained $j=(n-1)/2$.
For quantization of projections the well-known result is
valid\footnote{Because of non-complete understanding of the
quantization rules that time, in \cite{Born34}, \cite{Eps} and
\cite{Pauli} the erroneous value of the angular quantum number $j$
was ascribed as $j=n/2$, i.e. the quantum numbers $n_{1,2}$ took
the semi-integer values instead of integer and vice versa.}
\begin{equation} \fl
\frac{1}{2c}({\bi J}_1\cdot{\tilde{\bi B}}_1)=n_1\omega_1, \ \ \
\frac{1}{2c}({\bi J}_2\cdot{\tilde{\bi B}}_2)=n_2\omega_2, \ \ \
n_{1,2}=-j,-j+1,..., j-1,j . \label{2.1.7}
\end{equation}
The first correction to the energy $E_1$ is the perturbation
$\lambda V({\bi r},{\bi v})=-({\bi F}\cdot{\bi r})+({\bi
B}\cdot{\bi L})$ averaged over the Kepler period. Employing
(\ref{2.1.7}), it can be written in the form
\begin{equation}
E_1=\omega_1 n_1+\omega_2 n_2. \label{2.1.8}
\end{equation}
This result coincides with the quantum first-order correction
\cite{Pauli2}, \cite{DMO}.

In the case of crossed electric and magnetic fields
$\omega_1=\omega_2$. The first-order correction is degenerated
like in the case of Stark or Zeeman effects and an individual
state cannot be defined. This degeneracy is removed in the second
order of perturbation theory. This theory was developed in quantum
approach only \cite{Sol83}. However, a classical approach can be
developed in the same manner as in the case of the quadratic
Zeeman effect, which is presented in the next section.

\subsubsection{Quadratic Zeeman effect}

The problem of a hydrogen atom in a magnetic field has fundamental
importance. Attention to this problem significantly increased
after the discovery that energy splitting at avoided crossing
between adjacent manifolds decreases exponentially when the
principal quantum number $n$ increases \cite{Zim}.

In contrast to a hydrogen atom in an electric field the problem
with a magnetic field is not separable. For the sake of
definiteness, we choose the orientation of the homogeneous
magnetic field $\bi{B}$ along the $z$-axis. Since the Hamiltonian
of the hydrogen atom in a magnetic field ($\rho^2=x^2+y^2$,
$\omega=B/2c$ is the cyclotron frequency)
\begin{equation}
H=\frac{p^2}{2}-\frac{1}{r}+\frac{\omega^2 \rho^2}{2}+\omega L_z
\label{2.1}
\end{equation}
is invariant under rotation around the $z$-axis, the
$L_z$-component is conserved. Then, the motion along the azimuthal
angle $\varphi$ is separated out and the semiclassical
quantization condition along $\varphi$ gives
\begin{equation}
L_z=m,  \label{2.2}
\end{equation}
where $m=0,\pm 1,\pm 2, ...$ is the magnetic quantum number. After
the separation of the azimuthal angle $\varphi$ the problem is
reduced to a 2D non-separable one.

At ${\bi B}=0$ the electron moves on the Kepler elliptic orbit. In
this case, the angular momentum $\bi{L}$ and the Runge-Lenz vector
$\bi{A}$ are the two additional (to the energy) integrals of
motion. Under action of the weak magnetic field these integrals
start to change in time slowly, so that in the first order of
perturbation theory the combination ($\vartheta$ is the angle
between the Runge-Lenz vector $\bi{A}$ and the $z$-axis)
\begin{equation}
\Lambda=4A^2-5A_z^2=A^2(4-5\cos^2\vartheta),  \label{2.5}
\end{equation}
is conserved \cite{Sol81}. Taking into account that $0\leq A^2\leq 1$, the range of $%
\Lambda$ values is $-1\leq \Lambda^2\leq 4$. For $\Lambda=0$ the
Runge-Lenz vector lies on the surface of the double cone $\Omega$
specified by the condition $\cot \vartheta_0=2$ (see Fig.1). For
$\Lambda\neq 0$ all trajectories split up into two classes: the
trajectories with $\Lambda<0$ librate inside the double cone
$\Omega$ and trajectories with $\Lambda>0$ librate outside this
cone. Thus, all states are localized in two non-overlapping
domains. This unique property leads to the effect of the
exponential smallness of energy splitting at avoided crossings
between adjacent manifolds discovered by Zimmerman \textit{et al}
\cite{Zim}.
\begin{figure}
\begin{center}
\includegraphics*[width=10cm, height=6cm, angle=0]{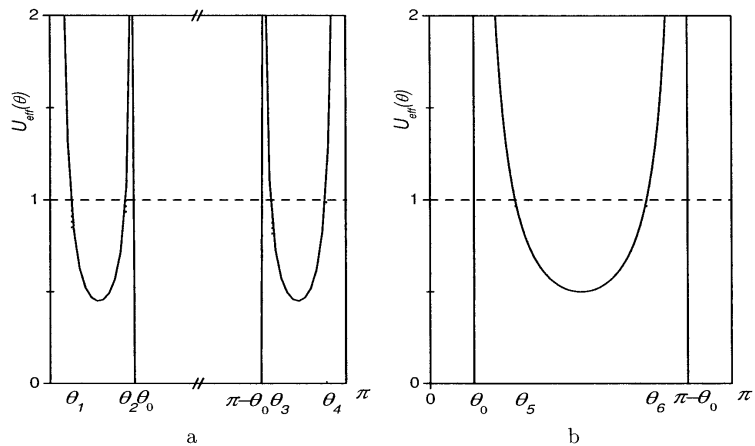}
\end{center}
\caption{ Effective potential $U_{eff}(\vartheta)$ as a function
of $\vartheta$ for $\Lambda=-0.5,\ \mu=0.1$ (a), and $\Lambda=2,\
\mu=0$ (b). } \label{fig:edin}
\end{figure}

The generalized momentum conjugated to the coordinate $\vartheta$
is the angular momentum component perpendicular to the plane
passing through the $z$-axis and the Runge-Lenz vector $\bi{A}$
($\mu=m/n$)
\begin{equation}
L_{\bot}(\vartheta)=n\sqrt{1+\frac{\Lambda}{1-5\sin^2\vartheta}- \frac{\mu^2%
}{\sin^2\vartheta}}=n\sqrt{1-U_{eff}(\vartheta)}.  \label{2.6}
\end{equation}
Fig.1 shows the effective potential $U_{eff}(\vartheta)$ for two
different cases: $\Lambda<0$ and $\Lambda>0$.

For negative values of $%
\Lambda$, the motion in intervals [$\vartheta_1,\vartheta_2$] and [$%
\vartheta_3,\vartheta_4$] is classically allowed (see Fig.1a), the
value of $\Lambda$ is double degenerate and the final expression
for the quantization rules has the form  \cite{Sol82}
\begin{eqnarray}
\int_{\vartheta_{1}}^{\vartheta_{2}}L_{\bot}(\vartheta)d\vartheta=\pi(s+1/2),
\nonumber \\
\int_{\vartheta_{3}}^{\vartheta_{4}}L_{\bot}(\vartheta)d\vartheta=\pi(s+1/2),
\qquad s=0,1,2,... \label{2.7}
\end{eqnarray}
Equations (\ref{2.7}) give two degenerate states which are
symmetric and
antisymmetric to the $xy$ plane. In the case of $m=0$ the turning points $%
\vartheta_1$ and $\vartheta_4$ do not occur. Instead, the two
singularities appear at $\vartheta=0,\pi$ due to the caustic at
the $z$-axis which is characterized by the same Morse index as an
ordinary turning point (see, e.g., \cite{LandM} \S 49). Therefore,
the quantization conditions in this case are formally obtained
from the quantization conditions (\ref{2.7}) by setting
$\vartheta_1=0$ and $\vartheta_4=\pi$. The analysis of the roots
of the function $L_{\bot}(\vartheta)$ shows that the states with
$\Lambda<0$ exist only when $m<n/\sqrt{5}$.

When $\Lambda>0$, the region of a classically allowed motion is the interval [$%
\vartheta_5,\vartheta_6$] (see Fig.1b). The $\Lambda$ values in
this case are nondegenerate and are determined from the
quantization condition  \cite{Sol82}

\begin{equation}
\int_{\vartheta_{5}}^{\vartheta_{6}}L_{\bot}(\vartheta)d\vartheta=%
\pi(k+1/2), \qquad k=0,1,2,...   \label{2.8}
\end{equation}
These states are localized outside the double cone $\Omega$ and their parity
with respect to the $xy$ plane is equal to $(-1)^k$.

In the first (with respect to $\omega^2$) order of perturbation
theory the quadratic Zeeman energy shifts are expressed in terms of the scaled value $%
\rho^2/n^2$ averaged over one period of the Kepler orbit

\begin{equation}
\varepsilon=\frac{<\rho^2>}{n^2}=\frac{1}{2}(1+\mu^2+\Lambda).  \label{2.9}
\end{equation}
The $\Lambda$ values are determined by the quantization conditions (\ref{2.7}%
) and (\ref{2.8}). Of greatest interest are the outmost levels in a given $%
\{nm\}$-manifold, since they are the first to undergo overlap in
the course of the approach to each other of two neighboring
manifolds with increasing strength of the magnetic field. These
energy levels correspond to the lowest levels in the effective
potential $U_{eff}(\vartheta)$ in the quantization conditions
(\ref{2.7}) and (\ref{2.8}). For these states the potential
$U_{eff}(\vartheta)$ can be approximated by the harmonic
oscillator and the scaled energy shift inside the cone $\Omega$ is
($\Lambda<0$)
\begin{equation}
\varepsilon_{osc.}=\sigma\sqrt{5+25\sigma^2}+\sqrt{5}\mu-5\sigma^2
\label{2.10}
\end{equation}
and outside the cone $\Omega$ ($\Lambda>0$)
\begin{equation}
\varepsilon_{osc.}=\frac{5}{2}-\kappa\sqrt{5+\frac{25}{16}\kappa^2-\mu^2}+%
\frac{5}{4}\kappa^2-\frac{3}{2}\mu^2 .  \label{2.11}
\end{equation}
where $\sigma=(2s+1)/n$ and $\kappa=(2k+1)/n$. In Table 1 the
scaled energy shifts (\ref{2.10}) and (\ref{2.11}) are compared
with the quantum calculations. The quantum results have been
obtained by diagonalisation of the energy matrix within the given
\{$nm$\}-subspace. The agreement for the position of levels is
very good for low-lying states. For higher states the agreement
becomes less satisfactory since the applicability of harmonic
oscillator approximation breaks down.

\begin{table}[tbp]
\caption{Comparison of the quantum results $\varepsilon_{q}$ with
the semiclassical approximation $\varepsilon_{osc.}$
(Eqs.(\ref{2.10}) and (\ref {2.11})).}
\begin{center}
\begin{tabular}{ll}
\begin{tabular}{c}
$n=40,\ m=0$ \\
\begin{tabular}{llllll}
\hline $s$ & $\varepsilon_{q}$ & $\varepsilon_{osc.}$ & $k$ &
$\varepsilon_{q}$ & $\varepsilon_{osc.}$ \\ \hline 0 & 0.055 &
0.053 & 0 & 2.45 & 2.44 \\ 1 & 0.159 & 0.142 & 4 & 2.03 & 2.06 \\
2 & 0.255 & 0.212 & 8 & 1.65 & 1.75 \\ 3 & 0.342 & 0.267 & 12 &
1.32 & 1.51 \\ \hline
\end{tabular}
\end{tabular}
&
\begin{tabular}{c}
$n=40,\ m=4$ \\
\begin{tabular}{llllll}
\hline $s$ & $\varepsilon_{q}$ & $\varepsilon_{osc.}$ & $k$ &
$\varepsilon_{q}$ & $\varepsilon_{osc.}$ \\ \hline 0 & 0.251 &
0.276 & 0 & 2.43 & 2.43 \\ 1 & 0.338 & 0.365 & 2 & 2.22 & 2.22 \\
2 & 0.415 & 0.436 & 4 & 2.01 & 2.04 \\ 3 & - & - & 6 & 1.82 & 1.88
\\ \hline
\end{tabular}
\end{tabular}
\end{tabular}
\end{center}
\end{table}
With the increase of the magnetic field the first avoided crossing
arises between the lowest-energy state $|1\rangle$ of the $n+1$
manifold and the highest-energy state $|2\rangle$ of the $n$
manifold. In the classical approach instead of avoided
crossing we obtain the exact crossing of energy curves since the first ($%
\Lambda\simeq-1$) and the second ($\Lambda\simeq4$) state are
located in two nonoverlapping regions of configuration space (see
Fig.1). The splitting is obtained in the quantum approach as the
matrix element of diamagnetic potential between the first
$|1\rangle$ and the second $|2\rangle$ state:
\begin{equation}
\Delta E=\langle 1|\omega \rho^2|2\rangle.  \label{2.13}
\end{equation}
In the first order of quantum perturbation theory the
Schr\"odinger equation is separable in elliptical-cylindrical
coordinates on a sphere in a four-dimensional momentum space
\cite{Sol81,Sol82,Her82}. Using uniform semiclassical
approximation for the wave functions of the states $|1\rangle$ and
$|2\rangle$ in this coordinate system (see, e.g., \cite{Groz84}),
the splitting is obtained in the form \cite{Sol82}
\begin{equation}
\Delta E\sim\exp\{-n\ln[(\sqrt{5}+2)(\sqrt{5}+1)/2]\}\simeq\exp(-1.92n),
\label{2.12}
\end{equation}
which coincides with the probability of the under-barrier penetration in the
effective potential $U_{eff}(\vartheta)$ with $\Lambda=-1$ in the interval $%
\vartheta\in[\vartheta_2,\vartheta_0]$ (see Fig.1a), and with
$\Lambda=4$ in the interval
$\vartheta\in[\vartheta_0,\vartheta_5]$ (see Fig.1b). Figure 2
demonstrates perfect agreement between this result and
experimental data \cite{Zim}.
\begin{figure}
\begin{center}
\includegraphics*[width=10cm, height=8cm, angle=0]{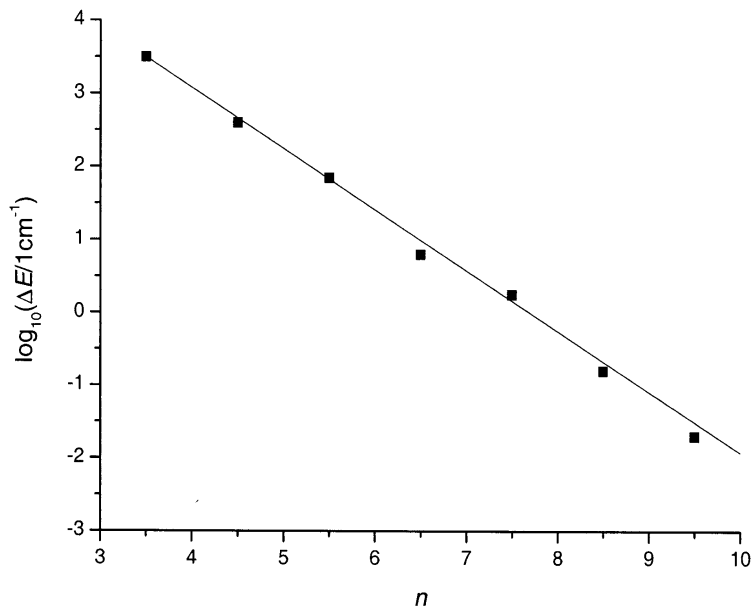}
\end{center}
\caption{Energy splitting between the lowest-energy state of the
$n+1$ manifold and the highest-energy state of the $n$ manifold as
a function of principal quantum number $n$: solid line - the
approximation (\ref{2.12}), solid squared - experimental data from
\cite{Zim}. } \label{fig:dva}
\end{figure}

\subsubsection{Helium atom; equivalent electrons}
The first attempts to develop a semiclassical perturbation theory
for a helium atom were made in the old Bohr theory (see, e.g.,
review \cite{TRR}). According to heuristic concepts accepted at
that time, only the simplest symmetric trajectories were
considered, which is wrong from the modern point of view . The
self-consistent perturbation theory was developed for equivalent
electrons (having the same principal quantum number $n$) with
total angular momentum equal to zero \cite{Sol85}.

A distinguishing feature of the classical perturbation theory for
equivalent electrons of a heliumlike system with the nuclear
charge $Z$ is the presence of the accidental degeneracy in the
unperturbed state\footnote{Accidental degeneracy means the
commensurability of the oscillation periods for two or several
coordinates, which takes place not always but at some initial
conditions.}. In this case, the perturbation series is the series
over {\it half-integer} power of the small parameter $\lambda=1/Z$
(see \cite{Poi} \S108). To construct this series, first of all,
the proper variables should be introduced. When the total angular
momentum is equal to zero, the trajectories of the two electrons
are on the same plane, regardless of the magnitude of the
electron-electron interaction, and their angular momenta are equal
but oppositely directed. In the zeroth order (without
electron-electron interaction) both electrons move along the
Kepler ellipses. The mutual orientation of the ellipses in the
plane is specified by an angle $\vartheta=\vartheta_1-\vartheta_2$
with $\vartheta_i$ being the angle between the Runge-Lenze vector
of the $i$-th electron and $z$-axis (see Fig.3). The position of
the $i$-th electron on the ellipse is determined by the Kepler
anomaly $\xi_i$ which is related to time by the relation
\begin{equation}
t-\tau_i=\frac{T_i}{2\pi}(\xi_i-\sqrt{1+2E_iL_i^2/Z^2}\sin{\xi_i})
\end{equation}
where $\tau_i$ is the instant of passage through the perihelion,
$E_i$ is the energy, $L_i$ is the angular momentum and $T_i$ is
the period of the $i$-th electron. Since the dependence on time of
the variables $\xi_1$ and $\xi_2$ is not independent, one of them
(for the sake of definiteness $\xi=\xi_1$) and the time delay
$\tau=\tau_1-\tau_2$ between the passage of the first and second
electrons through the perihelion should be used.

\begin{figure}
\begin{center}
\includegraphics*[width=10cm, height=3cm, angle=0]{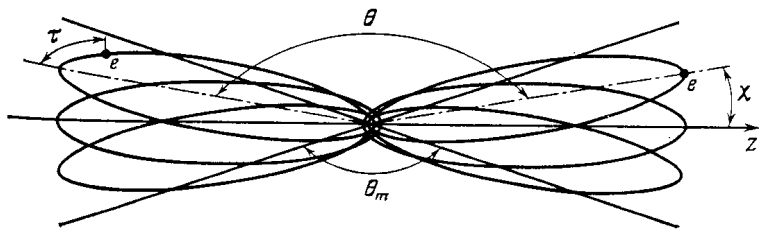}
\end{center}
\caption{Trajectories of two equivalent electrons for the case
$\chi_m=\pi/10$. From \cite{Sol85}.} \label{fig:three}
\end{figure}

In the planar case (${\bi L}\equiv{\bi L}_1=-{\bi L}_2$), the
motion of both electrons is described by eight Hamiltonian
equations. It follows from four of them that the quantities
($L_1-L_2$) and ($\vartheta_1+\vartheta_2$) are exactly conserved,
while ($E_1+E_2$) and ($\tau_1+\tau_2$) are conserved in the
first-order. A nontrivial role is played by the equations
\begin{eqnarray}
\frac{d\Delta E}{dt}=-\lambda\frac{\partial V}{\partial\tau}, \ \
\frac{d\tau}{dt}= \frac{3}{E_0}\Delta E,\label{2.1.3.2}
\\
\frac{dL}{dt}=-\lambda\frac{\partial V}{\partial\vartheta}, \ \
\frac{d\vartheta}{dt}=-\lambda\frac{\partial V}{\partial
L},\label{2.1.3.3}
\end{eqnarray}
where $E_0=-Z^2/2n$ is the unperturbed energy of one electron,
$\Delta E=(E_1-E_2)/2$ and $V(\xi,L,\vartheta,E_1,E_2,\tau)$ is
the electron-electron Coulomb interaction. The small parameter
$\lambda$ can be eliminated from equations (\ref{2.1.3.2}) and
(\ref{2.1.3.3}) by changing from $\Delta E$ to $\varepsilon=\Delta
E/\sqrt{\lambda}$ and introducing new 'times' $s_1=\sqrt{\lambda}
t$ and $s_2=\lambda t$ for the first and second pairs of the
equations, respectively. It follows hence that the rate of change
of $\tau$ and $\varepsilon$ is of the order of $\sqrt{\lambda}$,
and that of $L$ and $\vartheta$ of the order of $\lambda$, i.e.,
the angular momentum $L$ and mutual angle $\vartheta$ vary
infinitely slowly compared to time delay $\tau$ and scaled energy
$\varepsilon$ as $\lambda\to 0$. In addition, the energy transfer
$\Delta E$ is always small of the order of $\sqrt{\lambda}$,
therefore the change of the parameters $E_1$ and $E_2$, as well as
the periods  in the electron-electron interaction should therefore
be neglected, and these arguments of $V$ will hereafter be
omitted. Since the characteristic frequencies are different, the
motion along variables {$\varepsilon,\tau$} and {$L,\vartheta$} is
adiabatically separated.

Most sensitive to electron-electron interaction are the variables
$\varepsilon$ and $\tau$. They oscillate with period of the order
of $\lambda^{-1/2}$, during which $L$ and $\vartheta$ can be
regarded as constant. On the other hand, the oscillation over the
Kepler anomaly $\xi$ has a high frequency compared to
$\varepsilon$ and $\tau$. Replacing the interaction $V$ by its
averaged value over common period ($T\equiv T_1=T_2$)
\begin{equation}
{\cal
V}(\tau;L,\vartheta)=\frac{1}{T}\int_0^TV(\xi(t),L,\vartheta,\tau)dt
\label{2.1.3.4}
\end{equation}
we obtain one-dimensional problem in which $L$ and $\vartheta$
enter as parameters. The important feature of this one-dimensional
problem is that only the ground state of the potential
$\lambda{\cal V}(\tau;L,\vartheta)$ corresponds to the unperturbed
state and for quantization the corresponding Schr\"{o}dinger
equation should be used. Since $\Delta E$ and $\tau$ are
canonically conjugated variables, the Schr\"{o}dinger equation
reads
\begin{equation}
\Bigl[-\frac{3}{2E_0}\frac{d^2}{d\tau^2}+\lambda {\cal
V}(\tau;L,\vartheta)\Bigr]\psi(\tau;L,\vartheta)={\cal
E}(L,\tau)\psi(\tau;L,\vartheta) \label{2.1.3.5}
\end{equation}
with the periodic boundary condition:
$\psi(0;L,\vartheta)=\psi(T;L,\vartheta)$. For all excited states
of equation (\ref{2.1.3.5}) $\Delta E$ is finite at $\lambda=0$
and they do not go over into the states of the unperturbed
problem. In the first order of perturbation theory the wave
function of the ground state is constant, which is
$\psi_0=1/\sqrt{T}$. The first-order correction
\begin{equation}
{\cal E}(L,\vartheta)=\int_0^T\psi_0^2(\tau){\cal
V}(\tau;L,\vartheta)d\tau=\frac{1}{T}\int_0^T {\cal
V}(\tau;L,\vartheta)d\tau
\label{2.1.3.6}
\end{equation}
is additional integral of motion which is an effective Hamiltonian
in the angle $\vartheta$. After the substitution of
(\ref{2.1.3.4}) into (\ref{2.1.3.6}) this expression takes the
form
\begin{equation}
{\cal E}(L,\vartheta)=\frac{1}{T^2}\int_0^T \int_0^T
\frac{dt_1dt_2}{|{\bi r}_1(t_1)-{\bi r_2}(t_2)|},
\end{equation}
which coincides with classical averaging but for the nondegenerate
case. This leads to the important conclusion that from the point
of view of the quantization conditions there is no difference
between the nondegenerate and accidental degenerate motion.

\begin{figure}
\begin{center}
\includegraphics*[width=8cm, height=6cm, angle=0]{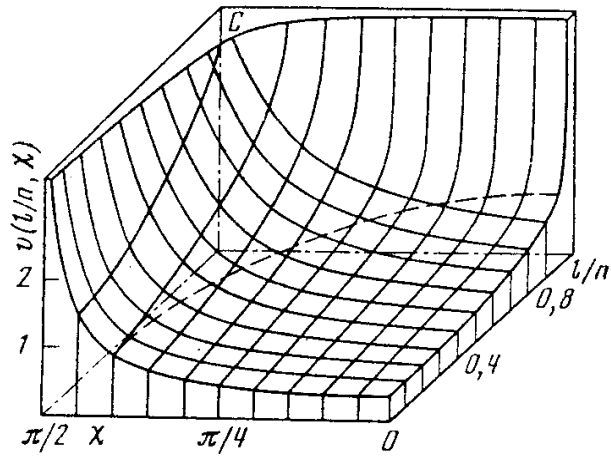}
\end{center}
\caption{Effective Hamiltonian as a function of $\nu$ and $\chi$.
The dashed line is the function $L(\chi)$ for $\chi_m=2\pi/5$ and
$q=0.61$. From \cite{Sol85}.} \label{fig:four}
\end{figure}

Using the explicit expression for the Kepler ellipses the scaling
with respect to the principal quantum number $n$ is obtained

\begin{equation}
{\cal E}(L,\vartheta)=\frac{Z}{n^2}v(\nu,\vartheta)
\label{2.1.3.9}
\end{equation}
where $\nu=L/n$. The numerically calculated effective Hamiltonian
$v(\chi,\nu)$ is shown in Fig.4.

The dependence of $L$ on $\vartheta$ is determined by the
condition that ${\cal E}(L,\vartheta)$ is a constant on the
trajectory, i.e.
\begin{equation}
v(\nu,\vartheta)=w \label{2.1.3.10}
\end{equation}
The constant $w$ is determined from the quantization condition,
which (in terms of the variable $\chi=(\pi-\vartheta)/2$) is the
same as considered for a hydrogen atom in a magnetic field at
$m=0$ (see Sec.2.1.1). Finally, the quantization condition takes
the scaled form
\begin{equation}
\int_0^{\chi_m}\nu(\chi)d\chi=\frac{\pi}{2}q \label{2.1.3.11}
\end{equation}
where $\chi_m$ is the turning point, $q=(2k+1)/2n$ and $k$ is a
new quantum number $(k=0,1,2,..., n-1)$. In the first order, the
correction $E^{(1)}$ to the unperturbed energy is equal to the
average electron-electron interaction

\begin{equation}
E^{(1)}=\frac{Z}{n^2}w(q). \label{2.1.3.12}
\end{equation}
Figure 5 shows a plot of $w(q)$ and also the corrections,
recalculated in accordance with the scaling rule (\ref{2.1.3.12}),
to the unperturbed energy in first-order quantum perturbation
theory for helium and obtained by exact calculation for He and
B$^{3+}$ \cite{Sol85}. As it is seen, the agreement improves with
increasing nuclear charge $Z$. The discrepancy between classical
and quantum perturbation theory becomes noticeable only at
$q>0.5$. It is attributable to the fact that as $q\to 1$ the
turning point $\chi_m$ approaches the singular point $\chi=\pi/2$
(see Fig.4). In this case, the quantization condition
(\ref{2.1.3.11}) must be replaced by the modified quantization
condition in which simultaneous account is taken of the turning
point and the singular point at $\chi=\pi/2$.

\begin{figure}
\begin{center}
\includegraphics*[width=6cm, height=6cm, angle=0]{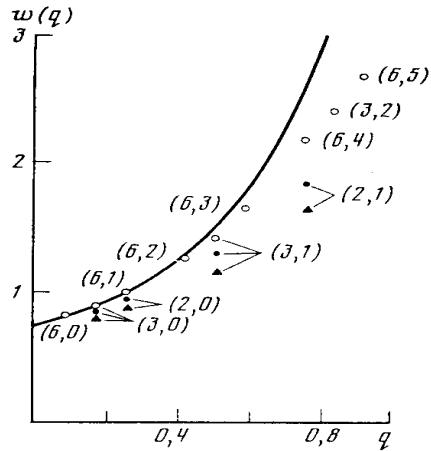}
\end{center}
\caption{Scaled correction to the unperturbed energy: solid line -
classical perturbation theory, open circles - quantum perturbation
theory, solid triangles - exact values for He, solid circles -
exact values for B$^{3+}$. The parentheses contain the state
quantum numbers ($n,k$). From \cite{Sol85}.} \label{fig:five}
\end{figure}

\setcounter{footnote}{0}
\subsection{Adiabatic switching method; hydrogen atom in external fields}

The extension of the semiclassical quantization to the
non-separable systems was proposed by Einstein in 1917 \cite{Ein},
who pointed out that the classical trajectory is Lagrangian
manifold, so if the classical trajectory forms a torus in phase
space, then in the $N$ dimensional problem only $N$ topologically
independent quantization contours on this torus exist, and the
actions along these contours are, according to the Liouville
theorem, invariant with respect to the deformation of the
contours. However, for a system with several degrees of freedom
the calculation requires an excessive amount of computer time,
mainly spent in rejecting unsuitable trajectories. To avoid this
problem, the adiabatic switching method was proposed \cite{Sol78}.
This method is based on the quantum Born-Fock adiabatic theorem
\cite{Born28} and the correspondence principle between quantum and
classical mechanics\footnote{Often, in the literature this method
is erroneously associated with the Ehrenfest adiabatic principle
\cite{Ehr}. The Ehrenfest principle was formulated for separable
problems only. Moreover, it was proved that in the non-separable
case it breaks down (see, e.g., \cite{Fermi}).}. The adiabatic
switching procedure is quite simple. To calculate energy spectrum
of non-separable Hamiltonian $H$, first of all, we need to guess a
reference Hamiltonian $H_0$ which is, on the one hand, solvable
(separable) and, on the other hand, has the same topology of
caustic as Hamiltonian $H$. Then one have to compute with the
classical equations of motion the development in time of the
initially quantized trajectory during a slow switching on the
interaction $V=H-H_0$ ($H(\lambda)=H_0+\lambda(t)V$). When the
interaction has been fully switched on ($\lambda=1$), the
quantized trajectory for Hamiltonian $H$ and corresponding
eigenvalue of energy are obtained, as well as, for all
intermediate values of switching parameter $\lambda$. It is
expected that the more slowly the interaction is switched on, the
more precisely the quantization conditions are satisfied. Of
course it is heuristic technics without proof but it works in many
cases; it is a simple and effective tool to calculate the energy
spectrum of involved multi-dimensional systems. The only thing
necessary to control is that the topology of caustics does not
change during the switching \cite{Sol78}.

In the case of hydrogen atom in crossed electric $\bi F$ and
magnetic $\bi B$ fields, the reference Hamiltonian $H_0$ is the
Hamiltonian of the hydrogen atom \cite{Gr82}. Due to the high
dynamical symmetry of the Coulomb interaction all trajectories in
configuration space are closed lines - ellipses and just one
quantization condition can be written which specifies principal
quantum number $n$ only. When external perturbation is introduced,
the trajectory fills up 3D domain in the configuration space and
missed quantization conditions can be formulated. The perturbation
theory for hydrogen atom in crossed electric and magnetic field is
presented in Sec.2.1.1. Thus, the initial conditions for the
quantized trajectory is determined by conditions (\ref{2.1.7}).
Next step is numerical calculation of classical equations for
electron -
\begin{equation}
\frac{d^2{\bi r}}{dt^2}+\frac{\bi r}{r^3}=-\lambda(t)\Bigl[{\bi
F}+\frac{1}{c}({\bi v}\times {\bi
B})\Bigr]+\frac{1}{c}\frac{d\lambda(t)}{dt}{\bi A} \label{2.3.2.1}
\end{equation}
- during the switching external fields from $\lambda=0$ to the
final values $\lambda=1$. In equation (\ref{2.3.2.1}) $\bi A$ is
the vector-potential (${\bi B}={\rm rot}{\bi A}$). The last term
on right-hand side of equation (\ref{2.3.2.1}) is the additional
force arising as a consequence of the time dependence of the
magnetic field. Although this force, being proportional to the
switching rate, disappears in the adiabatic limit, it cannot be
neglected. It ensures the adiabatic invariance, e.g., in the case
of a pure magnetic field ($F=0$). Obviously, if it were neglected,
the electron energy would be conserved exactly, that is wrong.
\begin{figure}
\begin{center}
\begin{tabular}{cc}
\includegraphics*[width=6.cm, height=6.cm, angle=0]{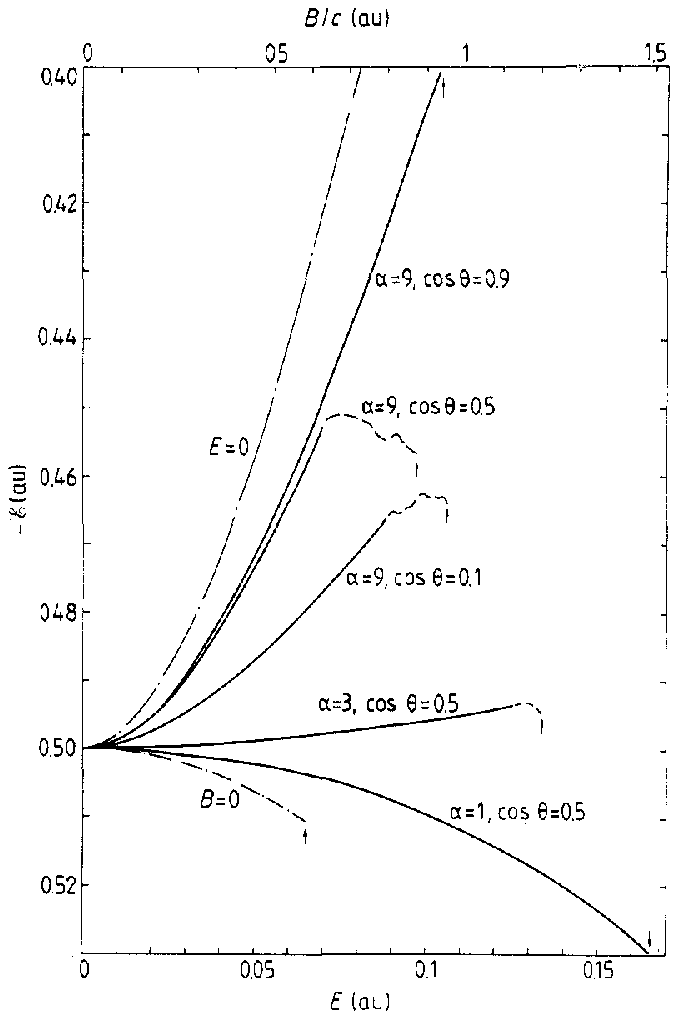} &
\includegraphics*[width=6.cm, height=6.cm, angle=0]{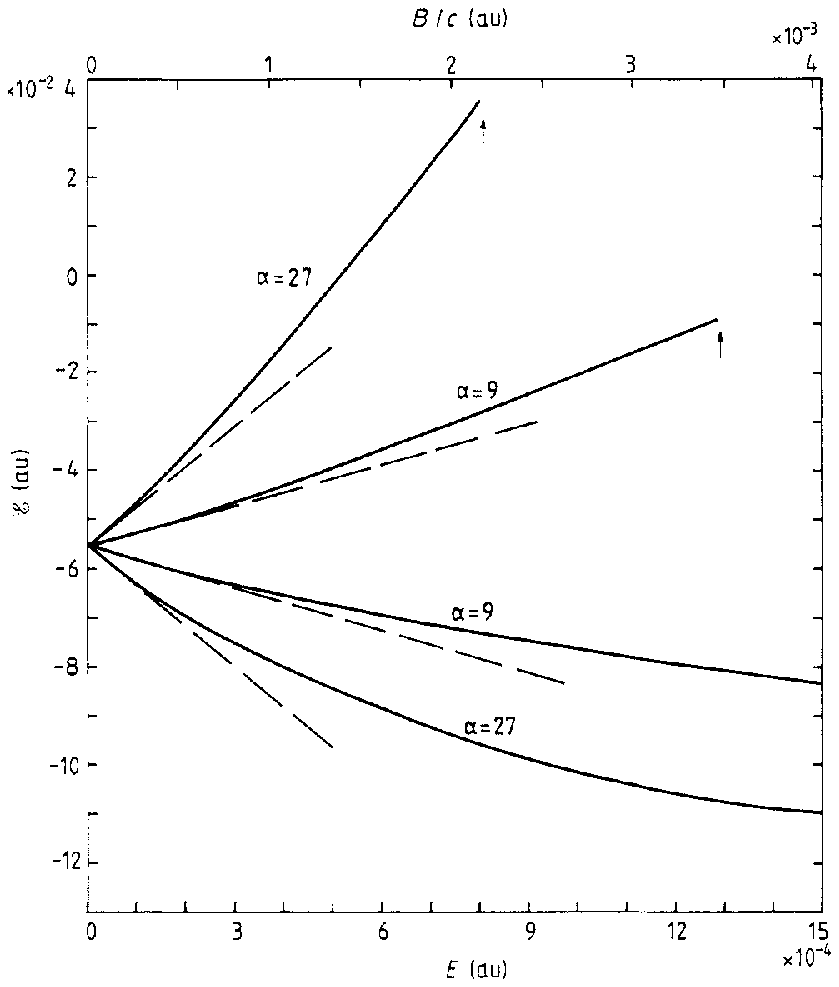}
\\
a) & b)
\end{tabular}
\end{center}
\caption{a) Ground state energy levels as functions of field
strength for various combinations of parameters $\alpha = B/cE$
and $\cos\theta$. Chain curves are the quantum results for the
cases of pure electric \cite{Sil78} and magnetic \cite{Cab72}
fields. Broken curves demonstrate the non-adiabatic evolutions of
the system. b) Energy levels of excited states,
$(n,n_1,n_2)=(3,-1,-1)$ (the levels shifted downwards) and
$(n,n_1,n_2) =(3,1,1)$ (the levels shifted upwards), as functions
of field strength with constant parameters $\alpha= 9, 27$ and
$\cos\theta=0.5$. Broken lines are the first-order perturbation
theory results (\ref{2.1.8}). The arrows indicate ionization in
both figures. From \cite{Gr82}. }
\end{figure}

Figure 6a shows the energy levels (full curves) of the ground
state as a function of field strengths, for different combinations
of parameters $\alpha=B/cF$ and the angle $\theta$ between $\bi F$
and $\bi B$. As a limit, quantum-mechanical results (chain curves)
for the ground-state energies of a hydrogen atom in a magnetic
field only ($E=0$) and in an electric field only ($B = 0$) are
shown. The curve labeled by $E = 0$ is the result \cite{Cab72},
obtained by diagonalising the energy matrix in an extensive basis
of hydrogen wavefunctions. The curve labeled by $B = 0$ is a
quantum fourth-order perturbation theory result (see, for example,
\cite{Sil78}). Figure 6b shows the results (full curves) for the
states defined by the weak-field quantum numbers $(n, n_1,n_2) =
(3, -1, -1)$ (the levels shifted downwards) and $(n,n_1,n_2) =(3,
1, 1)$ (the levels shifted upwards). The angle between the fields
has been kept constant ($\cos\theta = 0.5$) and the results are
presented for two different ratios of field strengths: $\alpha =
9$ and $\alpha = 27$. Unlike the ground-state case, here the
linear shift, given by (\ref{2.1.8}) (broken lines), dominates in
the limit of weak fields. The state (3,-1,-1) is bound stronger
and the ionization limit was not reached in the range of fields
shown in figure 6b. For the (3,1,1)-state the non-adiabatic
behaviour was found only very close to the ionization limits. In
Fig.6 electric field scale is common to all curves, while the
magnetic field scale corresponds to curves labeled by $\alpha=9$
for the ground state and $\alpha = 27$ for excited states. In all
cases, the switching rate used in calculations was $d\lambda(t)/dt
= 2.5 \times 10^{-5}$ au.

Again, adiabatic switching method has certain advantages with
respect to the  straightforward semiclassical quantization of
non-integrable systems. It is free of such {\it non-local}
problems as finding the caustics of the classical system and
searching for the initial conditions which correspond to quantized
trajectories (see, for example, \cite{N77}). In addition, this
method provides all intermediate energy values at $0\leq \lambda
\leq 1$.

\subsection{Spontaneous decay of excited states of a hydrogen atom}

A very interesting situation takes place in the problem of
spontaneous decay of excited states of a hydrogen atom. From the
classical point of view this process is due to the bremsstrahlung
(see, e.g., \cite{Af,Sp}). In the old Bohr quantum theory the
bremsstrahlung was the main obstacle to the description of the
stable atomic ground state within the framework of classical
mechanics since the accelerating electron should emit radiation
losing energy and finally should fall down onto the nucleus.
However, it does not contradict the evolution of the excited
states which are unstable.

According to classical electrodynamics, the rates of decrease of the energy $%
E(t)$ and the angular momentum $L(t)$ are described by the system
of equations (see, e.g., \cite{LandFT} \S 75)

\begin{table}[tbp]
\caption{Classical lifetime $\tau_{cl}$ (Eq.(\ref{2.4.2})) and
quantum lifetime $\tau_q$ \protect\cite{Wiese} for spontaneous
decay of excited states of a hydrogen atom (in nanoseconds).}
\begin{center}
\begin{tabular}{lllllllllllll}
\hline $nl$ & 2$p$ & 3$p$ & 4$p$ & 5$p$ & 6$p$ & 3$d$ & 4$d$ &
5$d$ & 6$d$ & 4$f $ & 5$f$ & $6f$ \\ \hline $\tau_{cl}$ & 1.68 &
5.66 & 13.4 & 26.2 & 45.3 & 15.7 & 37.3 & 72.8 & 126 & 73.1 & 143
& 247
\\ $\tau_q$ & 1.60 & 5.27 & 12.3 & 22.2 & 40.8 & 15.5 & 36.2 &
69.7 & 119 & 72.5 & 140 & 240 \\ \hline
\end{tabular}
\end{center}
\end{table}

\begin{eqnarray}
\frac{dE(t)}{dt}=-\frac{\{-2E(t)\}^{3/2}}{3c^3L(t)}\{3+2E(t)L^2(t)\},
\nonumber \\
\frac{dL(t)}{dt}=-\frac{2\{-2E(t)\}^{3/2}}{3c^3L^2(t)}.
\label{2.4.1}
\end{eqnarray}
The lifetime $\tau$ is defined by the expression
$P=const\times\exp(-t/\tau)$, where $P$ is the population of the
excited state. In classical mechanics there is no such concept,
but we can estimate it as the time $\tau_{cl}$ in which the
initial angular momentum decreases by unity ($\Delta L=1$), which
corresponds to the emission of one photon. In the leading order of
classical approach ($n,l \to \infty$) we can neglect the
dependence on time in the right-hand side of equations
(\ref{2.4.1}).
Then, after the substitution for the energy $%
E=-1/2n^2$ and angular momentum $L=(l+ 1/2)$ we obtain the
following estimate of lifetime in terms of the quantum numbers
$n,\ l$ \cite{Sp}
\begin{equation}  \label{2.4.2}
\tau_{cl}=\frac{3}{2}c^3n^3(l+1/2)^2=9.32\times
10^{-11}n^3(l+1/2)^2 \ \mathrm{sec}
\end{equation}
In Table 2 these results for several states are presented in
comparison with quantum calculations of total lifetimes $\tau_q$
of decay from the $nl$-state into all low-lying states
\cite{Wiese}. This table demonstrates amazing agreement even for
the low values of quantum numbers $n$ and $l$, though the lifetime
is not well-defined in classical theory.

The system of equations (\ref{2.4.1}) has an exact analytical
solution but, according to the correspondence principle, only the
leading classical order with respect to $n, l \to \infty$ has
physical meaning. The next terms compete with unknown quantum
corrections. Since the corrections of the same order caused by the
different sources usually compensate each other, we obtain worse
results taking them partly into account. It is the reason of bad
agreement obtained in \cite{Af} where classical and quantum orders
were mixed up.

Classical physics is deterministic, that is why the decay process
has one channel providing the total lifetime only, and there is no
way to analyze state-selective lifetime.

Generally, decay processes are not well-defined in quantum
mechanics. Here we can not separate the quantum system from
measurement. From the very beginning, it is formulated as a
exponential decreasing in time of the population of state which is
rather a classical treating of the problem. It is not clear how to
determine the width of the energy level in the degenerated case,
e.g., in the case of hydrogen atom in excited state at $n=2$. If
we add weak spherical perturbation, the degeneracy is moved and
correct wave functions are the spherical functions $|2s\rangle$
and $|2p\rangle$. Since the dipole matrix element between
$|2s\rangle$ and ground state $|1s\rangle$ is equal to zero,
$|2s\rangle $ is a meta-stable state having macroscopic lifetime -
$1.4$ sec. The second state $|2p\rangle$ has typical atomic
lifetime $1.6\times 10^{-9}$ sec. If we add weak electric field
the correct wave functions are characterized by
parabolic quantum numbers $n_1,\ n_2,\ m$ ($n=n_1+n_2+m+1$). At $n=2$ and $%
m=0$ there are two states $|nn_1n_2\rangle=|210\rangle$ and
$|201\rangle$ which have the same lifetime of order $2.3\times
10^{-9}$ sec. But what is the lifetime for the hydrogen without
any perturbation? There are huge difference between lifetime in
spherical and parabolic coordinates. Here we have an unsolvable
conflict with the superposition principle which is typical of
classical processes.

\subsection{Gutzwiller's approach; broad resonances}

The contribution of the unstable periodic orbit to the trace of
the Green function is determined by the Gutzwiller formula
\cite{Gutz}
\begin{equation}
g(E)\sim -\frac{iT(E)}{2\hbar}\sum_{n=1}^{\infty}\frac{\exp\{in[S(E)/\hbar-%
\lambda\pi/2]\}}{\sinh[nw(E)/2]}  \label{3.1}
\end{equation}
where $S(E)$, $w(E)$, $T(E)$ and $\lambda$ are the action, the
instability exponent, the period and the number of focal points
during one period, respectively. After the expansion of the
denominator, according to
$[\sinh(x)]^{-1}=2e^{-x}\sum_{k=0}^{\infty}e^{-2kx}$, and
summation of the geometric series over $n$ one can see that the
response function (\ref{3.1}) has poles at the complex energies
$E_{ks}$ whenever
\begin{equation}
S(E_{ks})=\hbar\lambda\pi/2-i\hbar w(E_{ks})\left(k+\frac{1}{2}%
\right)+2s\pi\hbar,  \ \ k,s=0,1,2,...  \label{3.2}
\end{equation}
Equation (\ref{3.2}) is a transcendental equation with respect to
$E_{ks}$. This approach is not well-defined \cite{Sol10}, but, in
the case of the shortest period associated with unstable periodic
orbit in separable system, the poles of (\ref{3.1}) can be
interpreted as manifestation of the resonances in the continuum.
The width of these resonances is the linear function of $\hbar$ in
spite of the under-barrier resonances whose width is exponentially
small with respect to $\hbar \to 0$ ( $\propto e^{-2|S|/\hbar}$,
where $S$ is under-barrier action). The Gutzwiller's approach was
applied to the realistic systems - the scattering of electron on
Coulomb potential in the presence of magnetic \cite{Win87,Du87}
and electric fields \cite{Win89,Gr91}, and scattering on
two-Coulomb-center potential \cite{Gr91}.

\begin{table}[tbp]
\caption{Real parts $E_{G}$ and widths $\Gamma_{G}$ (in atomic
units) for a sequence of resonances with parabolic quantum numbers
$(n_1,n_2,m)$ at field strength $F=8$ kV/cm, as obtained from
Gutzwiller's approach (\ref{3.56}). $E_q$ and $\Gamma_q$ are the
results of the full quantum calculations \cite{Kol86}.}
\begin{center}
\begin{tabular}{lllllll}
\hline $(n_1,n_2,m)$
&(23,0,0)&(23,1,0)&(23,0,1)&(24,0,0)&(24,1,0)&(24,0,1)
\\ \hline
   $E_{G} \times 10^4$ & 1.899 & 1.774 & 2.597 & 3.360 & 3.251 &
   4.021
\\ $E_{q} \times 10^4$   & 1.949  & 2.039 & 2.698 & 3.382 & 3.433 &
4.090
\\ $\Gamma_{G} \times 10^4$  & 0.499 & 1.480 & 1.145 & 0.632 & 1.887 &
1.355
\\ $\Gamma_q \times 10^4$      & 0.524 & 1.681 & 1.188 & 0.638 & 1.955 & 1.369 \\

 \hline
\end{tabular}
\end{center}
\end{table}

As an example let us consider resonancees in elastic scattering of
electron on the Coulomb centre in the uniform electric field $\bi
F$ \cite{Gr91}. The Hamiltonian is separable in parabolic
coordinates $u,v$ and $\varphi$ ($x^2+y^2=u^2v^2$, $2z=u^2-v^2$
and $u,v\geq 0$). After introducing a new time-like variable
$\tau$ defined by the relation $d\tau=dt/(u^2+v^2)$, the problem
is reduced to two separated one-dimensional systems:
\begin{eqnarray}
H_1=\frac{p_u^2}{2}+\frac{m^2\hbar^2}{2u^2}-Eu^2+\frac{1}{2}Fu^4=2\beta,
\label{3.3}
\\
H_2=\frac{p_v^2}{2}+\frac{m^2\hbar^2}{2v^2}-Ev^2-\frac{1}{2}Fv^4=2(1-\beta),
\label{3.4}
\end{eqnarray}
where $\beta$ is a separation constant. At positive energy $E$ the
motion in the coordinate $u$ is bounded whereas in the coordinate
$v$ it is unbounded. The system (\ref{3.3}),(\ref{3.4}) has
hyperbolic fixed point whose location in the $v-p_v$ space at
energy $E=E_{ks}$ is given (up to first order in $\hbar$) by
\begin{equation}
v^*=\sqrt{|m|\hbar}(2E_{ks})^{-1/4}e^{i\pi/4}, \ \ p_{v}^*=0, \ \
\beta^*=1+i|m|\hbar\sqrt{E_{ks}/2} \label{3.5}
\end{equation}
This fixed point corresponds to unstable state since in its
vicinity the Hamiltonian $H_2$ has a form of turned oscillator
($\delta p_v=p_v-p_v^*,\ \delta v=v-v^*$)
\begin{equation}
\delta^2 H_2=\frac{1}{2}(\delta p_v)^2 -4E_{ks}(\delta v)^2
\label{3.55}.
\end{equation}
In the first order with respect to $\hbar\to 0$ the corresponding
action and instability exponent are
\begin{eqnarray}
S(E_{ks})=2\int_0^{u_0}\sqrt{2E_{ks}u^2+4-Fu^4}du- |m|\pi
\hbar-\frac{i}{2}|m|\hbar w(E_{ks}), \label{3.6}
\\
w(E_{ks})=4\sqrt{2E_{ks}}\int_0^{u_0}\frac{du}{\sqrt{2E_{ks}u^2+4-Fu^4}},
\label{3.7}
\end{eqnarray}
where $u_0$ is the external turning point. The substitution
(\ref{3.6}), (\ref{3.7}) into (\ref{3.2}) gives the transcendental
equation for complex energy $E_{n_1n_2}$ ($s=n_1$, $k=n_2$,
$\lambda=2$)
\begin{equation}
S(E_{n_1n_2})=(2n_1+1)\pi\hbar+i\hbar\Bigl(n_2+\frac{1}{2}\Bigr)w(E_{n_1n_2})
\label{3.56}.
\end{equation}
which is, up to the first order of $\hbar$, equivalent with
equations (15)-(17) in Ref.\cite{Gr88} obtained in quantum
approach by the comparison equation method. Table 3 demonstrates a
good agreement between Gutzwiller's approach (\ref{3.56}) and full
quantum calculations \cite{Kol86} for the series of resonances
which lie close to the real $E$-axis. They all correspond to large
values of parabolic quantum number $n_1$ and small values of $n_2$
and $m$. The widths $\Gamma_G$ are comparable with the differences
of the positions $E_G$ of the adjacent resonances. In this sense
they can be referred to as "broad resonances". They explain, e.g.,
the experimentally observed resonant structure of photoionization
cross section of the hydrogen atom near the ionization limit in
strong electric fields \cite{Kol86,Glab}.

This problem was considered by Wintgen too \cite{Win89}. His
result, although qualitatively good, does not give correct values
for the complex poles of expression (\ref{3.1}) in the case $m=0$.
The approach presented in \cite{Gr91} is based on the complex
periodic orbits in the $m\neq 0$ case and gives the correct result
for $m=0$ as $m\to 0$ limit.

The same approach with the same result was developed for broad
resonances in the scattering of electron on two-Coulomb-center
potential \cite{Gr91}.

\subsection{Poincar\'e section; resonances in helium atom}

The Poincar$\acute{\mathrm{e}}$ section is a powerful tool to
study the irregular motion in 2D non-separable problems because of
the visualization of a dynamical system on one plot. In this case,
the phase space is 4D with two coordinates $(x,y)$ and two momenta
$(X,Y)$. At fixed energy $E$ one of these variables can be
expressed from the equation $H(x,y,X,Y)=E$ as a function of the
energy $E$ and the remained variables. Let it be, for instance,
the second momentum $Y(x,y,X,E)$. Now if we plot on the plane
$\{x,y\}$ the values of variables $x(t)=x_i$ and $X(t)=X_i$ each
time $t_i$ when the coordinate $y(t)$ takes some value $y_0$ then,
in the general case, the distribution of points ($x_i,X_i$) in
this plane is irregular. But if an additional integral of motion
(may be approximate one) exists - $\Lambda(x,y,X,Y)$, then from
the transcendental equation
$\Lambda(x,y_0,X,Y(x,y_0,X,E))=\lambda$ we obtain the functional
dependence of $x$ on $X$ and constants $y_0$, $E$ and $\lambda$:
$x(X;y_0,E,\lambda)$, where $\lambda$ is the value of the second
integral of motion. As $t\to \infty$ ($i\to \infty$) the points
($x_i,X_i$) from one trajectory fulfill this line. Calculation of
many trajectories gives the Poincar$\acute{\mathrm{e}}$ section of
the system, which visualizes the information about regular and
irregular regions of motion. In the regular region there can exist
two types of fixed points, elliptic and hyperbolic, associated
with stable and unstable periodic orbits, respectively.

\begin{figure}
\begin{center}
\begin{tabular}{cc}
\includegraphics*[width=5.cm, height=5cm, angle=0]{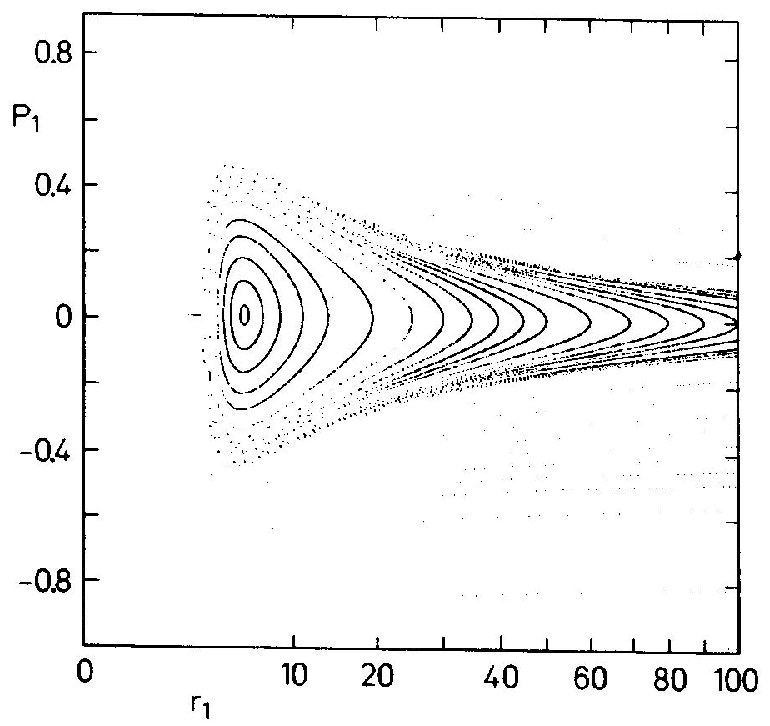} &
\includegraphics*[width=5.cm, height=5cm, angle=0]{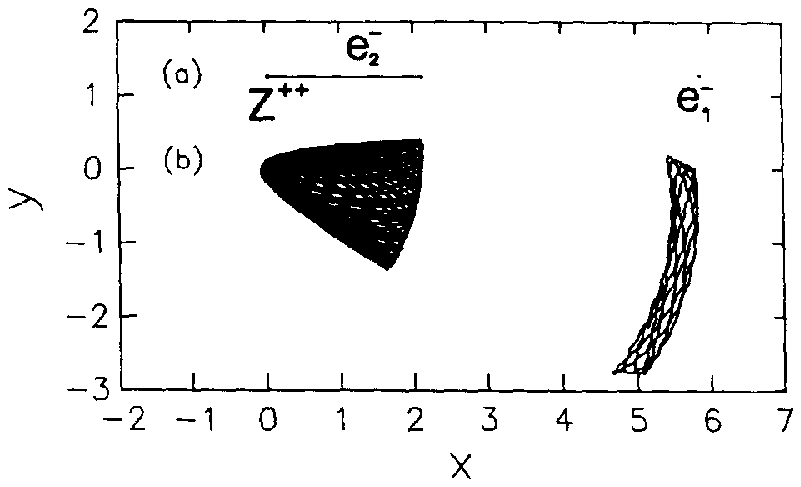}
\\
a) & b)
\end{tabular}
\end{center}
\caption{a) The Poincar$\acute{\mathrm{e}}$ section for collinear
configurations of the electrons. The coordinate $r_1$ and the
momentum $p_1$ of the outer electron is monitored whenever the
inner electron approaches the nucleus ($r_2=0$). b) The periodic
trajectory corresponding to the elliptic fixed point on the left
panel (a) and a nonperiodic but regular trajectory in its
neighbourhood (b). From \cite{Rich92}. }
\end{figure}
Consider a collinear arrangement of a nucleus $Z$ and of two
electrons, both being on the same side of the nucleus. It is a
two-dimensional problem (position of inner and outer electrons on
the axis), where we can use the Poincar$\acute{\mathrm{e}}$
section which is shown for helium in Figure 7a. In this section,
the phase space position ($r_1,\ p_1$) of the outer electron is
monitored each time the inner electron approaches the nucleus
($r_2=0$). From Figure 7 one can see the elliptic fixed point at
$p_1=0$ and $r_1\approx 7$, which corresponds to the two-electron
periodic orbit (Fig.7b upper trajectory). Such a motion is a
coherent oscillation of both electrons but with large difference
in their positions and velocities. The outer electron appears to
stay nearly frozen at large distance. This localization of outer
electron is amazing since the total charge of the core (the
nucleus and inner electron) is attractive. The existence of such
states is explained by the resonant interchange of the energy
between two electrons, i.e. it is a purely dynamical effect - at
the conventional analysis we should expect that outer electron
should fall down onto the nucleus since the total charge of
nucleus and inner electron is equal to -1. Thus, it is very
surprising that these classical configurations are extremely
stable.

Semiclassical quantization near the periodic orbit is based on the
Gutzwiller's formula (\ref{3.1}), but, since the orbit is stable,
the instability exponent must be replaced by the stability
parameter $\gamma=-iw$. The final expression of the quantized
classical action of the tori surrounding the periodic orbit is
given by \cite{Rich92}
\begin{equation}
S_{nkl}=s+1/2+2(k+1/2)\gamma_1+(l+1/2)\gamma_2, \ \ \ \ \
s,k,l=0,1,2,... \label{2.3.1.1}
\end{equation}
where $\gamma_1=0.46164$ and $\gamma_2=0.06765$ the winding
numbers, which describe the behaviour of nearby trajectories. The
quantum numbers $s,\ k$ and $l$ reflect the approximate
separability in the local coordinates parallel and perpendicular
to the periodic orbit; $s$ is the quantum number along the
periodic orbit, $k$ represents bending degree of freedom and $l$
corresponds to the perpendicular degree but preserving
collinearity. Using the scaler property $S=S_{sc}/\sqrt(-2E)$, the
quantized energy levels are obtained
\begin{equation}
E_{nkl}=-\frac{S_{sc}^2}{[s+1/2+2(k+1/2)\gamma_1+(l+1/2)\gamma_2]^2},
 \label{2.3.1.2}
\end{equation}
where $S_{sc}=1.4915$ is the (scaled) action of the periodic orbit
for helium.

\begin{table}[tbp]
\caption{Semiclassical energies $E_{scl}$ from expression
(\ref{2.3.1.2}), quantum energies $E_q$ and widths $\Gamma_q/2$
for the states of the series $(s,0,l))$. From \cite{Rich92}.}
\begin{center}
\begin{tabular}{lllllll}
\hline $(s,k,l)$ &(4,0,0)& (7,0,0)&(4,0,1)&(7,0,1)&(4,0,2)&(7,0,2)
\\ \hline $E_{scl} \times 10^2$ & -8.91  & -3.48 & -8.68 & -3.42 & -8.45 & -3.36
\\ $E_{q} \times 10^2$ & -8.96  & -3.48 & -8.76 & -3.43 & -8.61 & -3.39
\\ $\Gamma_q/2 \times 10^6$  & \ 2.02 & \  0.37 & \ 6.60 & \ 0.61 & \ 0.79 & \ 0.70\\
 \hline
\end{tabular}
\end{center}
\end{table}

Table 4 shows the semiclassical energies (\ref{2.3.1.2}) in
comparison with full quantum calculation of energies $E_q$. In the
table the width of the states $\Gamma_q$ with respect to the
autoionization is also presented. One can see extremely small
values of the widths for this series of asymmetric double excited
states.

\section{Ion-atom and electron-atom collisions}

\subsection{Electron-impact detachment of negative ion near the
threshold} The ionization of a negative ion by an electron impact
draws much attention because of a huge cross section. This
reaction is very important, for instance, in understanding of
stellar atmospheres.

First of all we need a classical description of the negative ion.
The classical model for interaction of electron with neutral core
of negative ion is based on the same physical assumptions as the
zero-range potential model in the quantum theory \cite{DO}, i.e.
the potential well supporting one weakly bound $s$-state is
spherical, narrow, and deep. The electron is oscillating along the
diameter between the opposite turning points $d$ and $-d$. The
frequency of this oscillation $\omega$ is the large parameter of
the classical model. The bound $s$-state is represented by the
ensemble of such trajectories uniformly distributed in all
directions.

Now let us consider the detachment of the negative ion by an
electron as a projectile. Near the threshold, because of the
Coulomb repulsion between a negative ion and a projectile, there
is another large parameter - the distance of the closest approach
of the projectile to the negative ion $R_0$. Since $R_0$ is large,
there is no interaction between the projectile and neutral core.
Then, the equation of motion for the projectile is
\begin{equation}
\frac{d^2{\bi r}_1}{dt^2}=\frac{{\bi r}_1-{\bi r}_2}{|{\bi
r}_1-{\bi r}_2|^3} \label{3.2.1}
\end{equation}
where ${\bi r}_1$ is the position vector of the projectile and
${\bi r}_2$ is the position vector of the bound electron. Since
$r_2\leq d$, $r_1\geq R_0$ and $R_0\gg d$, we can expand the
right-hand side of equation (\ref{3.2.1}) over the multipole
series. For our purposes, it is sufficient to consider the dipole
approximation
\begin{equation}
\frac{d^2{\bi r}_1}{dt^2}=\frac{{\bi r}_1}{r_1^3}-\frac{{\bi
r}_2}{r_1^3}+\frac{3({\bi r}_1\cdot{\bi r}_2){\bi r}_1}{r_1^5}
\label{3.2.2}
\end{equation}
The first term on the right-hand side of equation (\ref{3.2.2})
describes the interaction with the fixed Coulomb center, but the
second and third represent the rapidly oscillating force with
frequency $\omega$. Kapitsa developed an approximation method for
this situation (see, e.g., \cite{LandM}, \S 30). In this method,
the position vector of the projectile splits into two terms ${\bi
r}_1=\bi{R}+\bi{r}$, where $\bi{R}$ is a smoothly-varying function
and $\bi{r}$ is the rapidly oscillating part having zeroth mean
value over a period $2\pi/\omega$. In the leading order with
respect to $\omega \to \infty$ this method leads to the following
results:
\begin{eqnarray}
\frac{1}{2}\dot{R}^2+\frac{Eb^2}{R^2}+\frac{1}{R}=E, \nonumber
\\
{\bi r}=\frac{1}{\omega^2}\Bigl[\frac{{\bi r}_2}{R^3}-3\frac{({\bi
r}_2\cdot{\bi R}){\bi R}}{R^3}\Bigr], \label{3.2.3}
\end{eqnarray}
where $b$ is the impact parameter and $E$ is the initial energy of
the projectile. Ionization occurs when the work done by the
projectile $A$ becomes equal to the binding energy of the negative
ion $\epsilon$:
\begin{equation}
A\equiv\int \frac{({\bi r}_1-{{\bi r}_2})\cdot d{\bi r}_2}{|{\bi
r}_1-{\bi r}_2|^3}= \epsilon \label{3.2.4}
\end{equation}
It is easily verified with the aid of (\ref{3.2.3}) that the work
done by a projectile is an oscillating function whose mean value
is zero, and the amplitude of the oscillation increases in the
course of the collision. Consequently, the energy necessary for
ionization is not accumulated gradually in the collision process,
but is transferred in a quarter of the period while the bound
electron moves from the nucleus in the direction opposite to that
of the projectile.  The orientation of each trajectory in the
ensemble is characterized by spherical angles $\vartheta$,
$\varphi$ with respect to the vector ${\bi R}(t)$. The classical
trajectories, for which escape is possible, are then restricted to
a cone $\Omega(t)$ with $\vartheta\leq\vartheta_m(t)$. The size of
the cone is obtained from (\ref{3.2.4}):
\begin{equation}
\frac{d\cos \vartheta_m(t)}{R^2(t)}=\epsilon. \label{3.2.5}
\end{equation}
The detailed analysis shows  that the cone corresponding to the
distance of the closest approach $\Omega_0$ absorbs  all previous
cones $\Omega(t)$. Therefore, the probability of ionization at a
given impact parameter $b$ is
\begin{equation}
W(b)=\frac{1}{2\pi}\int_{\Omega_0}\sin \vartheta d \vartheta
d\varphi=1-\frac{E_{th}^2}{4E^2}(1+\sqrt{1+4E^2b^2})^2
\label{3.2.6}
\end{equation}
where $E_{th}=\sqrt{\epsilon/d}$ is the {\it classical} threshold
of detachment. The normalization factor in equation (\ref{3.2.6})
is taken as $1/2\pi$ because the opposite directions represent the
same trajectory.  The total detachment cross section is
\cite{Sol77}
\begin{equation}
\sigma=2\pi\int_0^{b_m}
W(b)bdb=\frac{\pi}{2E_{th}^2E^4}(E-E_{th})^2\Bigl(E^2+\frac{2}{3}EE_{th}+\frac{1}{3}E_{th}^2\Bigr),
\label{3.2.10}
\end{equation}
where $b_m=\sqrt{E(E-E_{th})}/(E_{th}E)$. The classical threshold
energy $E_{th}$ differs from the quantum threshold energy
$\epsilon$, because, in the classical approach, the projectile
needs extra energy to overcome the Coulomb repulsion and to reach
the distance where ionization occurs. For $\epsilon<E<E_{th}$ the
detachment mechanism is purely quantum, it is an the underbarrier
penetration.

Figure 8 shows the cross section (\ref{3.2.10}) for
electron-impact detachment of H$^-$ (or D$^-$) in comparison with
experimental data. The parameters for H$^-$ were taken from
\cite{DO}: $d=2.7$ and $\epsilon=0.0278$. The classical results
are in good agreement with experimental data \cite{Wal} and follow
the shape of experimental data \cite{And} in the threshold region.
However, they are shifted approximately by 2 eV with respect to
the latter. This shift is equal to the difference between
classical (2.76 eV) and quantum (0.76 eV) thresholds.

It is not clear how to reproduce the classical mechanism of
detachment in quantum mechanics. The detachment cross section in
the threshold region depends critically on the interaction between
the projectile and the rapidly oscillating bound electron. In
quantum mechanics, it is difficult to take this interaction into
account rigorously, whereas in classical approach this can be done
simply and naturally. In \cite{OT} the authors tried to modify the
classical approach. However, the proposed "corrections" are beyond
the selfconsistent asymptotic approach. For instance, it was taken
into account that not all electrons, satisfying the condition
$\vartheta<\vartheta_m$, have time to leave the negative ion
during collision, and, at the same time, the under barrier
penetration is considered, in fact, as an {\it instant} process.

\begin{figure}
\begin{center}
\includegraphics*[width=10cm, height=8cm, angle=0]{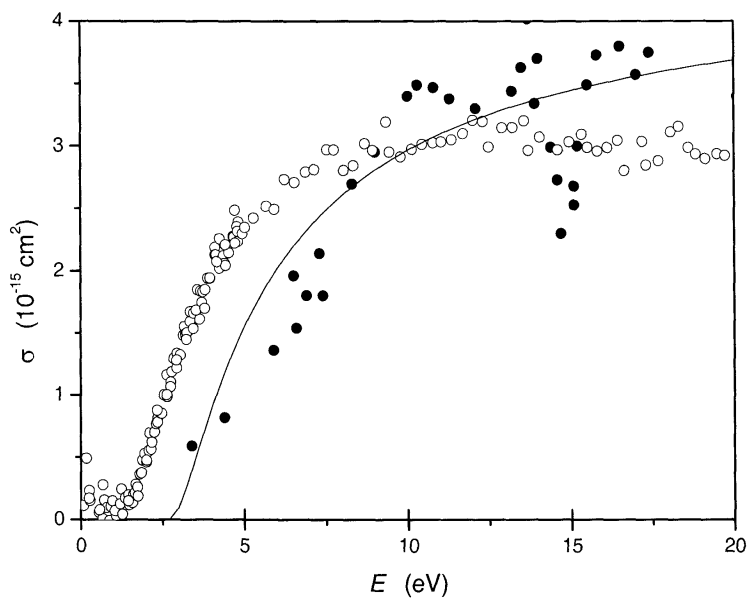}
\end{center}
\caption{The detachment cross section of H$^-$ (or D$^-$) as a
function of electron impact energy. The full curve is the
classical cross section (\ref{3.2.10}), the open circles are
experimental data from \cite{Wal} and the solid circles are
experimental data from \cite{And}. } \label{fig:tri}
\end{figure}

\subsection{Non-adiabatic transitions via hidden crossings}

In slow atomic collisions the inelastic transitions between
electronic adiabatic states occur at the place of the closest
approach of adiabatic potential curves - avoided
crossings\footnote{According to the von Neumann-Wigner theorem
\cite{NW}, the exact crossing of two adiabatic potential curves of
the same symmetry is forbidden.}. Initially in the application of
the adiabatic approach the narrow avoided crossings due to the
under-barrier resonant interaction of two electronic states
located on different nuclei were used. Later \cite{Sol81b} the new
mechanism of nonadiabatic transitions via the so-called 'hidden
crossings' were discovered. Hidden crossings happen when an
electronic energy level touches the top of an effective potential.
In the classical description the full dimensional electronic
trajectory collapses into an unstable periodic orbit at this
place. The transitions via hidden crossings are dominant in atomic
collisions. They provide a complete description of inelastic
processes in atomic collisions. The classical/semiclassical theory
of hidden crossings was developed in \cite{Sol86}. In detail it is
discussed in the review paper \cite{Sol05}.

\subsection{Binary encounter approximation}

For inelastic transitions of atomic electron in the charged
projectile-neutral atom collisions, the binary encounter
approximation can be applied. This approximation is based on the
assumption that inelastic cross sections can be obtained from the
two-body (projectile-atomic electron) {\it elastic} cross section
after averaging over the momentum distribution of atomic electron
in the initial atomic state. This scheme can be used in both
quantum and classical approach.

Initially, the classical binary encounter approximation was
developed for fast collisions (see, e.g., review paper \cite{Per}
and references therein). In \cite{Iv95}, the restriction on impact
velocity $v_p>>v_e$, where ${\bi v}_p$ and ${\bi v}_e$ are the
initial projectile and atomic electron velocities, was removed.
The most interesting case is the case when the projectile is an
electron. In this case, the electro-electron velocity transfer
from the projectile to the atomic electron is given by
\begin{equation}
\Delta {\bi v}_e=\frac{2(v^3{\bi b}-2{\bi v})}{{v^4b}^2+4},
\label{3.3.13}
\end{equation}
where ${\bi v}={\bi v}_p-{\bi v}_e$ is the electron-electron
impact velocity and $\bi b$ is a vector of impact parameter (${\bi
v}\bot{\bi b}$). Then the kinetic energy transfer is
\begin{equation}
\Delta E({\bi b})=\frac{2v_pv_ev^2b\sin \vartheta \cos
\chi+2v^2+4({\bi v}\cdot{\bi v}_e)}{v^4b^2+4}, \label{3.3.2}
\end{equation}
where $\vartheta$ is the angle between ${\bi v}_p$ and ${\bi
v}_e$, $\chi$ is the angle between vector of ${\bi b}$ and the
plane in which vectors ${\bi v}_p$ and ${\bi v}_e$ lie. The
boundary of the region of $\bi b$ values which lead to an energy
transfer greater than $\varepsilon$ is specified by the condition
$\Delta E({\bi b})=\varepsilon$, and the differential cross
section with respect to velocities of projectile ${\bi v}_p$ and
atomic electron ${\bi v}_e$ for the transfer of an energy $\Delta
E({\bi b})\geq\varepsilon$ is
\begin{equation}
\Sigma({\bi v}_p,{\bi v}_e,\varepsilon)=\int \Theta(\Delta E({\bi
b})-\varepsilon)d^2{\bi
b}=\frac{\pi}{v^4\varepsilon^2}({v'}_p^2{v'}_e^2-v_p^2 v_e^2\cos^2
\vartheta), \label{3.3.3}
\end{equation}
where $\Theta(x)$ is the step-wise Heaviside function, and
${v'}_p=\sqrt{v_p^2-2\varepsilon}$,
${v'}_e=\sqrt{v_e^2+2\varepsilon}$ are the projectile and atomic
electron velocities in the final state. Assuming the isotropic
distribution over ${\bi v}_e$ in the initial atomic state the
averaged cross section is obtained as
\begin{equation} \fl
\Sigma(v_p,v_e,\varepsilon)=
 \frac{\pi}{8\varepsilon^2}
  \cases {
 \frac{v_p^2+v_e^2}{v_pv_e}\ln{\frac{(v_p+v_e)^2}{(v_p-v_e)^2}}
-2\frac{({v'}_p^2-{v'}_e^2)^2}{({v}_p^2-{v}_e^2)^2}-2 & $v_p \geq
{v'}_e$, \cr \frac{v_p^2+v_e^2}{{v'}_p{v'}_e}
\ln{\frac{(v_p^2+v_e^2+2{v'}_p{v'}_e)}{(v_p^2+v_e^2-2{v'}_p{v'}_e)}}
-4 & $v_p < {v'}_e$. \cr } \label{3.3.4}
\end{equation}
The final expression for the cross section
$\sigma(v_p,\varepsilon)$ is found by taking an average of
(\ref{3.3.4}) over the distribution of electron in the atom
$q({\bi r})$. For the spherical symmetric atomic potential $V(r)$
the distribution in the ${nl}$-subshell has the form \cite{Iv93}
\begin{equation}
q_{nl}({\bi r})=\frac{1}{2\pi r^2 T_{nl} p_{nl}(r)}, \label{3.3.5}
\end{equation}
where $p_{nl}(r)= \sqrt{2[E_{nl}-V(r)-L^2/2r^2]}$ is the radial
momentum of atomic electron in the initial state $E_{nl}$,
$L=\hbar(l+1/2)$ and $T_{nl}$ is the period of oscillation over
$r$. Then the cross section is expressed as an integral between
the turning points $r_{1,2}$
\begin{equation}
\sigma_{nl}(v_p,\varepsilon)=\frac{2}{T_{nl}}\int_{r_1}^{r_2}
\Sigma(v_p,\sqrt{2[E_{nl}-V(r)]},\varepsilon)\frac{d
r}{p_{nl}(r)}. \label{3.3.6}
\end{equation}
For the fast collision ($v_p\to \infty$), the cross section does
not depend on the mass of projectile and has a general asymptote
\begin{equation}
\sigma_{nl}(v_p,\varepsilon)=\frac{2\pi
Z^2}{3v_p^2\varepsilon^2}[3\varepsilon+2(E_{nl}-\tilde{V}_{nl})],
\label{3.3.7}
\end{equation}
where $\tilde{V}_{nl}$ is the average potential energy (for
Coulomb potential $\tilde{V}_{nl}=2E_{nl}$) and $Z$ is the charge
of the projectile. At $\varepsilon=|E_{nl}|$ expressions
(\ref{3.3.6}),(\ref{3.3.7}) give the ionization cross section
$\sigma_{ion}(v_p)$.

To calculate the cross section for excitation to a state with a
principal quantum number $n'$, $\varepsilon$ in (\ref{3.3.5})
should be presented in the form $\varepsilon=E_{n'}-E_{nl}$ and
then after differentiation of (\ref{3.3.5}) with respect to $n'$
the cross section of excitation is obtained \cite{Iv95}
\begin{eqnarray}
\sigma_{nl\to
n'}(v_p)=\Bigl|\frac{\partial\sigma_{nl}(v_p,\varepsilon)}{\partial\varepsilon}
\frac{dE_{n'}}{dn'}\Bigr|_{\varepsilon=E_{n'}-E_{nl}}= \nonumber
\\
=\frac{4\pi}{T_{n'}T_{nl}}\Biggl|\int_{r_1}^{r_2}\frac{\partial\Sigma(v_p,\sqrt{2[E_{nl}-V(r)]},
\varepsilon)}{\partial\varepsilon}\frac{d
r}{p_{nl}(r)}\Biggr|_{\varepsilon=E_{n'}-E_{nl}},
 \label{3.3.8}
\end{eqnarray}
where $T_{n'}$ is the period of oscillation over $r$ in the final
state $n'$. In (\ref{3.3.8}) the fact was used  that in the
semiclassical approximation the derivative of the energy with
respect to the quantum number is equal to the classical frequency:
$dE_{n'}/dn'=2\pi/T'$ which follows straightforwardly from the
Bohr-Sommerfeld quantization condition.

\begin{figure}
\begin{center}
\begin{tabular}{cc}
\includegraphics*[width=6cm, height=5cm, angle=0]{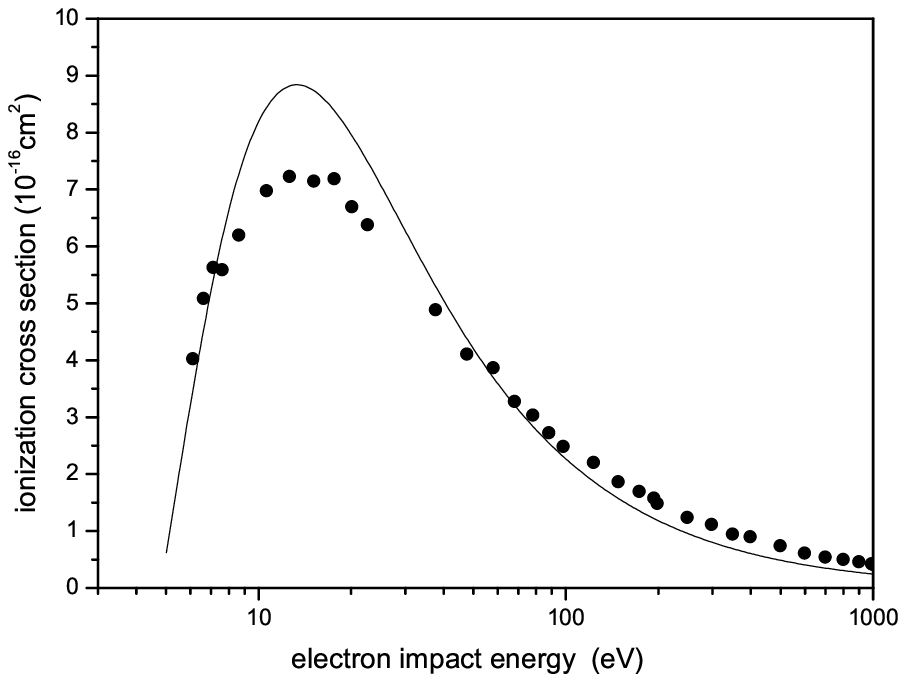} &
\includegraphics*[width=6cm, height=5cm, angle=0]{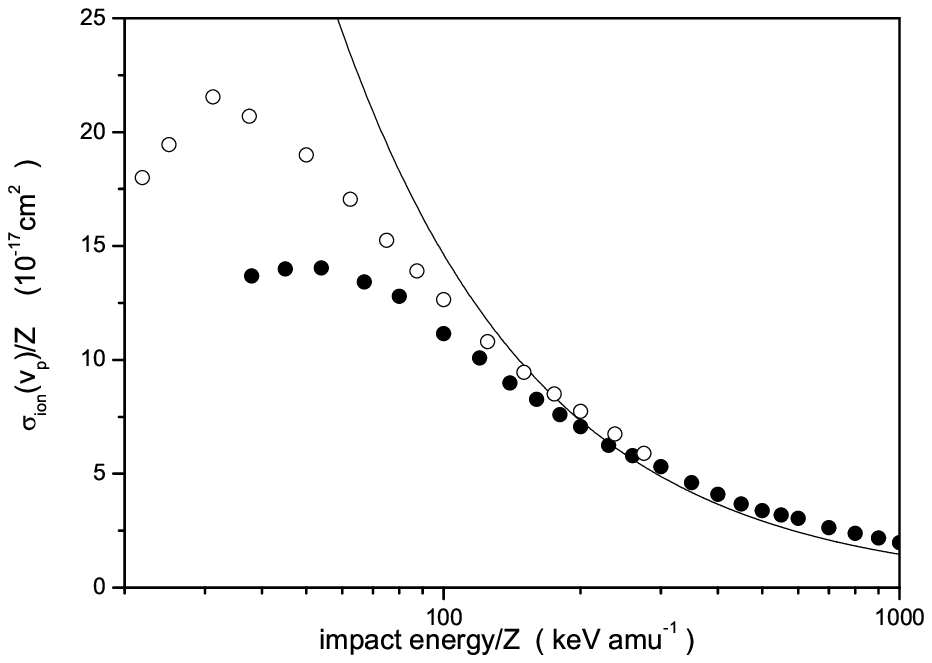}
\\
a) & b)
\end{tabular}
\end{center}
\caption{a) The ionization cross section of He($^3S$) metastable
as a function of electron impact energy. The full curve is the
classical cross section (\ref{3.3.6}) and the solid circles are
experimental data \cite{Dixon}. b) The ionization cross section of
hydrogen atom by proton and He$^{2+}$ as a function of scaled
impact energy. The full curve is the asymptote (\ref{3.3.7}), the
solid and open circles are the experimental data \cite{Shah} for
proton and He$^{2+}$ as a projectile.}
\end{figure}

Figure 9a shows the cross section for electron-impact ionization
of helium in the meta-stable state He(2$^3S$). It can be seen that
classical expressions give good agreement with experimental
results in the whole range of impact energy. In Fig.9b the
comparison of asymptote (\ref{3.3.7}) with experimental data for
ionization of hydrogen atom by proton and He$^{2+}$ is presented.
The discrepancy at very high impact velocity in both figures is
explained by logarithmic-like dependence of the cross section
($\sigma\propto \ln v_p/v_p^2$) due to the contribution from
individual {\it dipole} transition. However, according to
Ehrenfest's theorem, the classical description deals with the
wavepackets. In order to obtain a wavepacket, which imitates the
motion of a classical particle, one needs to perform integration
over the quantum numbers in a narrow range. As a result of this
averaging, the logarithmic energy dependence disappears. It
explains the absence of the logarithmic energy term in the
classical cross sections.

To calculate the state selective excitation $nl\to n'l'$, the
angular-momentum transfer is needed. The square of the angular
momentum in the final state can be written
\begin{equation}
L'^2=[{\bi r}\times({\bi p}+\Delta{\bi
p})]^2=L^2+r^2(2p_{\perp}\Delta p_{\perp}\cos \alpha+\Delta
p_{\perp}^2), \label{3.3.9}
\end{equation}
where ${\bi p}_\perp$ and $\Delta{\bi p}_{\perp}$ are the
projections of ${\bi p}$ and $\Delta{\bi p}$ onto the plane
perpendicular to $\bi r$, and $\alpha$ is the angle between ${\bi
p}_\perp$ and $\Delta{\bi p}_{\perp}$. Since for transition from
the $(nl)$-subshell the initial distribution $q_{nl}({\bi r})$ is
isotropic, the distribution with respect to $L'$ for fixed values
of $\bi r$ and $\bi b$ is
\begin{equation}
\lambda({\bi r},{\bi
b},L')=\frac{1}{2\pi}\Bigl|\frac{d\alpha}{dL'}\Bigr|=\frac{2L'}{\pi
r^2\sqrt{(\mu_2^2-\Delta p^2_{\perp})(\Delta p_{\perp}^2-\mu_1^2)}
}, \label{3.3.10}
\end{equation}
where $\mu_{1,2}=(L'\pm L)/r$. After averaging of (\ref{3.3.10})
over the impact parameter $\bi b$ and the initial distribution
$q_{nl}({\bi r})$ the final result is obtained as a single radial
integral over the classically allowed region which is common to
the initial and final states. The detailed analysis of this cross
section can be found in \cite{Iv93,Iv95}. In the case $n,n'\gg
\Delta n$ and $L,L'\gg \Delta L$ ($\Delta n=|n'-n|,\ \Delta
L=|L'-L|$) in high-energy limit the cross section has the form
\begin{equation}
\sigma_{nl \to n'l'}(v_p)=\frac{32n^3(n^2-L^2)^{3/2}}{9\pi
v_p^2(n\Delta n^2+n\Delta L^2-2L\Delta n \Delta L)^2}
\label{3.3.11}
\end{equation}
This expression reveals some important features. For fixed values
of $n$ and $L$, the cross section sharply decreases with
increasing $\Delta n$ and $\Delta L$, as $(\Delta n)^{-4}$ and
$(\Delta L)^{-4}$.

\section{Classical representation in quantum mechanics}

At first sight, the classical representation is impossible in
quantum theory, for instance, because of tunneling phenomenon,
which is, at first sight, incompatible with the classical
approach. However, if we consider a particle scattered by a
parabolic barrier, we find that, on the one hand, there is an
effect of sub-barrier penetration and, on the other hand, the
time-dependent Green function, which carries full information
about the quantum system, is exactly expressed in terms of the
classical action along classically {\it allowed} trajectories
only. Thus, in principle, the tunneling effect could be explained
in terms of classical trajectories.

\subsection{Abel transform to the classical representation}

For the sake of transparency we consider a symmetric potential
$V(x)=V(-x)$, where $V(0)=0$ and $V(x)$ increases monotonically to
infinity on the semiaxis $x>0$. To introduce the classical
representation the linear equation for the quantum probability
density $\rho_n(x)=\psi_n^2(x)$ is needed , which is obtained from
the Schr\"{o}dinger equation \cite{Sol93}
\begin{equation}
-\frac{\hbar^2}{4m}\frac{d^3\rho_n(x)}{dx^3}+2[V(x)-E_n]\frac{d\rho_n(x)}{dx}+\frac{dV(x)}{dx}\rho_n(x)=0.
\label{4.1.1}
\end{equation}
This third-order differential equation has three linearly
independent solutions - two of these are related to the
Schr\"{o}dinger equation and the third to the solution of the
Milne equation having the meaning of the quantum wavelength (see
Sec.5).

Now let us represent the quantum state $n$ as an ensemble of
classical states in the potential $V(x)$ distributed in the energy
$\epsilon$ with a certain probability density $\phi_n(\epsilon)$
\begin{equation}
\rho_n(x)=\int_{V(x)}^{\infty}q(\epsilon,x)\phi_n(\epsilon)
d\epsilon, \label{4.1.2}
\end{equation}
where
\begin{equation}
q(\epsilon,x)=\frac{\sqrt{2m}}{T(\epsilon)\sqrt{\epsilon-V(x)}}
\end{equation}
is the $x$-distribution for the classical state at given energy
$\epsilon$ and $T(\epsilon)$ is a classical period. A remarkable
property of the representation (\ref{4.1.2}) is its reciprocity.
This becomes evident if instead of $x$ the potential $V$ is taken
as an independent variable. Such a change of variable is standard
in classical mechanics and reduces the conversion of expression
(\ref{4.1.2}) to the well-known Abel problem, whence
\begin{equation}
\phi_n(\epsilon)=-\frac{T(\epsilon)}{\pi\sqrt{2m}}\int_{\epsilon}^{\infty}\frac{1}
{\sqrt{V-\epsilon}}\frac{d\rho_n(x(V))}{dV}dV, \label{4.1.3}
\end{equation}
where $x(V)$ is the inverse of function $V(x)$.

The transforms (\ref{4.1.2}), (\ref{4.1.3}) are an analogue of the
Fourier transform connecting configuration and momentum
representations and it is called 'a classical representation',
since the kernel $q(\epsilon,x)$ is the classical probability
density and $\phi(\epsilon)$ has the meaning of the energy
distribution in the classical ensemble.

In the classical representation, the Schr\"{o}dinger equation
(\ref{4.1.1}) transforms into an equation for scaled energy
distribution
$\tilde{\phi}_n(\epsilon)=\phi_n(\epsilon)/T(\epsilon)$:
\begin{equation}
(\epsilon-E_n)\tilde{\phi}_n(\epsilon)=\frac{\hbar^2}{
m}\int_{\epsilon}^{\infty}
Q(\mu,\epsilon)\frac{d^3\tilde{\phi}_n(\mu)}{d\mu^3}d\mu,
\label{4.1.4}
\end{equation}
with the kernel
\begin{equation}
Q(\mu,\epsilon)=\frac{1}{15\pi}\int_{x_1}^{x_2}\frac{1}{\sqrt{V(x)-\epsilon}}\frac{d^3(\mu-V(x))^{5/2}}
{dx^3}dx,\label{4.1.5}
\end{equation}
where $x_1$ and $x_2$ are the turning points defined by the
conditions: $V(x_1)=\epsilon$ and $V(x_2)=\mu$. The discrete
spectrum $E_n$ is obtained from the boundary condition:
$\tilde{\phi}_n(\epsilon)\to 0$ as $\epsilon \to \infty$. Equation
(\ref{4.1.4}) looks like a balance equation for 'virtual'
transitions between the classical states with the energies
$\epsilon$ and $\mu$. It is very interesting to extract the
probability of such transitions from equation (\ref{4.1.4}).

The energy distribution $\phi_n(\epsilon)$ obeys the obvious
normalization condition
\begin{equation}
\int_0^{\infty}\phi_n(\epsilon)d\epsilon=1
\end{equation}
In addition to this, the general condition holds
\begin{equation}
\int_0^{\infty}\epsilon\phi_n(\epsilon)d\epsilon=E_n,
\end{equation}
which is not trivial and means that the average energy of
classical ensemble is equal to the eigen-energy of quantum state
$|n\rangle$.

\subsection{Harmonic oscillator}
Let us consider the harmonic-oscillator problem
\begin{equation}
V(x)=\frac{1}{2}m\omega^2x^2
\end{equation}
as an example in which the classical representation is employed.
In this case the kernel can be calculated explicitly:
\begin{equation}
Q(\mu,\epsilon)=\frac{1}{4}m\omega^2(\mu-2\epsilon).
\end{equation}
Substituting this into (\ref{4.1.4}) yields an equation for
harmonic oscillator in classical representation
\begin{equation}
\frac{\hbar^2\omega^2}{4}\Bigl[\epsilon
\frac{d^2\phi_n(\epsilon)}{d\epsilon^2}+\frac{d\phi_n(\epsilon)}{d\epsilon}\Bigr]+(E_n-\epsilon)\phi_n(\epsilon)=0,
\end{equation}
which formally coincides with the radial Schr\"{o}dinger equation
for the Sturmian basis functions of the hydrogen atom with the
negative 'angular quantum number' $l=-1/2$. In the Sturmian
approach the energy is fixed and the charge is quantized. The
energy $E_n$ plays the role of the 'charge' in this 'hydrogen
Schr\"odinger equation', its spectrum is $E_n=\hbar\omega(n+1/2)$,
and the corresponding energy distribution reads
\begin{equation}
\phi_n(\epsilon)=\frac{2}{\hbar\omega}e^{-2\epsilon/\hbar\omega}L_n(4\epsilon/\hbar\omega),
\label{4.1.9}
\end{equation} where $L_n(x)$ is a Laguerre
polynomial.

\subsection{Inelastic transitions in the Feynman model}

As practical application of the classical representation, the
problem of description of the initial and final states in
classical trajectory Monte Carlo (CTMC) method \cite{AP} in the
theory of atomic collisions is the first candidate. In this method
the initial state of the electron is specified as an ensemble of
classical states with the energy equal to the energy $E_n$ of the
corresponding atomic state. Then the evolution of this ensemble is
calculated according to the classical equations of motion. The
main difficulty in this approach is to extract the contribution of
different quantum states in the final ensemble as $t \to \infty$,
since the final energies of classical trajectories assume
continuous values that do not coincide with the atomic spectrum.
On the other hand, the classical representation (\ref{4.1.3})
provides the exact description of quantum states in terms of an
ensemble of classical trajectories. This fact can be used to
modify the CTMC method in the treatment of the initial and final
states. Of course, the solution of the dynamical problem remains
classical, i.e. approximate. To illustrate such an approach, let
us consider inelastic transitions in the Feynman model \cite{Fey}
with the nonstationary potential
\begin{equation}
V(x,t)=\frac{1}{2}m\omega^2x^2+\alpha(t)x,
\end{equation}
where $\alpha(t)$ is the strength of a homogeneous external field
that is time-dependent and tends to zero as $t \to \pm \infty$. In
this case, the general solution of the classical equation of
motion can be obtained explicitly
\begin{equation}
x(t)=\frac{1}{m\omega}\int_{-\infty}^t\alpha(t')\sin \omega
(t-t')dt'+ \frac{\sqrt{2\epsilon}}{\omega}\sin{\omega(t-\tau)},
\label{4.1.11}
\end{equation}
where $\epsilon$ and $\tau$ are the initial energy and phase. As
$t\to +\infty$ this solution describes harmonic oscillation with
an energy
\begin{equation}
\mu=\epsilon+\nu+2\sqrt{\epsilon\nu}\cos{\tau},
\end{equation}
where
\begin{equation}
\nu=\frac{1}{2m^2}\Bigl|\int_{-\infty}^{+\infty}\alpha(t)e^{i\omega
t}dt\Bigr|^2.
\end{equation}
Since the distribution over the initial phase is uniform, the
energy distribution in the final state at initial energy
$\epsilon$ has the form
\begin{equation}
p(\mu,\epsilon)=\frac{1}{\pi}\frac{d\tau}{d\mu}=
\frac{\theta(\mu-\mu_1)\theta(\mu_2-\mu)}{\pi\sqrt{\theta(\mu-\mu_1)\theta(\mu_2-\mu)}},
\end{equation}
where $\mu_1=(\sqrt{\epsilon}-\sqrt{\nu})^2$,
$\mu_2=(\sqrt{\epsilon}+\sqrt{\nu})^2$ and $\theta(x)$ is the
unite step-function. Then the probability of an inelastic
transition from the state $|n\rangle$ to the state $|k\rangle$ is
\cite{Sol93}
\begin{equation}
P_{nk}(\gamma)=\int_0^{\infty}d\mu\phi_k(\mu)\int_{\epsilon_1}^{\epsilon_2}d\epsilon
p(\mu,\epsilon)\phi_n(\epsilon)
=L_n^{k-n}(\gamma)L_k^{n-k}(\gamma))e^{\gamma}, \label{4.1.15}
\end{equation} where
$\phi_k(\mu)$ and $\phi_n(\epsilon)$ are the energy distributions
of the initial and final states defined by (\ref{4.1.9}),
$\epsilon_1=(\sqrt{\mu}-\sqrt{\nu})^2$,
$\epsilon_2=(\sqrt{\mu}+\sqrt{\nu})^2$, $\gamma=\nu/\hbar\omega$
and $L_p^q(x)$ is a generalized Laguerre polynomial. The
probability (\ref{4.1.15}) coincides with the exact quantum
expression (see \cite{Fey}), although the classical solution
(\ref{4.1.11}) is approximate from a quantum point of view.

\subsection{Semiclassical approach}

The rough classical approximation is obtained if we neglect the
right-hand side in equation (\ref{4.1.4}). Then
$\phi_n(\epsilon)=\delta(\epsilon-E_n)$ and from (\ref{4.1.2}) the
density of probability takes the form
\begin{equation}
\rho_n(x)=\frac{\sqrt{2m}}{T(E_n)\sqrt{E_n-V(x)}}. \label{4.4.1}
\end{equation}
However, the solution $\tilde{\phi}_n(\epsilon)$ as well as
$\rho_n(x)$ has essential singularity as $\hbar \to 0$:
\begin{equation}
\tilde{\phi}_n(\epsilon)\propto
\sin\Bigl(\frac{1}{\hbar}W(\epsilon)\Bigr). \label{4.4.1'}
\end{equation}
To construct semiclassical approximation, the integral in the
right-hand side of equation (\ref{4.1.4}) has to be integrated by
parts twice
\begin{eqnarray}
(\epsilon-E_n)\tilde{\phi}_n(\epsilon)=\frac{\hbar^2}{
8m}\Bigl[\frac{dV(x)}{dx}\Bigl\vert_{V=\epsilon}\Bigr]^2\frac{d^2\tilde{\phi}_n(\epsilon)}{d\epsilon^2}+
\nonumber
\\
+\frac{\hbar^2}{m}\frac{d^2V(x)}{dx^2}\Bigl\vert_{V=\epsilon}\frac{d\tilde{\phi}_n(\epsilon)}{d\epsilon}+
\frac{\hbar^2}{m}\int_{\epsilon}^{\infty} \frac{d^2
Q(\mu,\epsilon)}{d\mu^2}\frac{d\tilde{\phi}_n(\mu)}{d\mu}d\mu.
\label{4.4.2}
\end{eqnarray}
Here the coefficients in front of the second and first derivatives
of $\tilde{\phi}_n(\epsilon)$ are obtained from definition of
$Q(\mu,\epsilon)$ (\ref{4.1.5}). Since $\tilde{\phi}_n(\epsilon)$
has essential singularity with respect to $\hbar$, in zero order
only the first term in the right-hand side should be retained.
Then $W(\epsilon)$ takes the form
\begin{equation}
W(\epsilon)=2\int^\epsilon\sqrt{2m(E_n-\epsilon')}\Bigl[\frac{dV(x)}{dx}\Bigl\vert_{V=\epsilon'}\Bigr]^{-1}d\epsilon'.
\label{4.4.3}
\end{equation}
After substitution (\ref{4.4.1'}) with $W(\epsilon)$ (\ref{4.4.3})
into expression (\ref{4.1.2}), the density of probability can be
estimated by the expression
\begin{equation}
\rho(x)\propto
\int^{\infty}_{V(x)}q(\epsilon,x)\sin\Bigl(\frac{2}{\hbar}\int^\epsilon\sqrt{2m(E_n-\epsilon')}
\Bigl[\frac{dV(x)}{dx}\Bigl\vert_{V=\epsilon'}\Bigr]^{-1}d\epsilon'\Bigr)d\epsilon.
\label{4.4.4}
\end{equation}
Semiclassical expansion of $\rho(x)$ is generated by consecutive
integration by parts in (\ref{4.4.4}). In the leading order
\begin{eqnarray}
\rho(x)\propto
\cos\Bigl(\frac{2}{\hbar}\int^x\sqrt{2m(E_n-V(x'))}dx'\Bigr)=
\nonumber
\\
=2\cos^2\Bigl(\frac{1}{\hbar}\int^x\sqrt{2m(E_n-V(x'))}dx'/\hbar\Bigr)-1.
\label{4.4.5}
\end{eqnarray}
The last term, -1, has to be omitted because it is beyond the
semiclassical series, and the result coincides with the well-known
semiclassical expansion of $\rho(x)$ in the leading order.

\subsection{Scattering problem; tunneling phenomenon}

In quantum mechanics, the scattering problem is out of Hilbert
space and we have no well-defined {\it physical} stationary states
because of normalization problem. It is a very serious defect of
the theory. Below the boundary of the continuum, the aim of the
theory is the search for stable (or stationary) states which exist
at discrete energy values only. Above the boundary of the
continuum the classical motion of particle is, in principle, not
stable (the particle comes from infinity and then goes to infinity
as $t \to \infty$), and we cannot clearly formulate a physical
problem like in the case of bound states. But, nevertheless, if we
accept the main idea of classical representation - the quantum
state is reproduced by an ensemble of classical states - then we
can interpret, for example, the tunneling phenomenon as
penetration through a barrier of part of this ensemble with energy
of trajectories above the top of a barrier.

\section{Semiclassical series}

\subsection{Higher orders of semiclassical series for quantum wavelength}

The semiclassical approximation for the one-dimensional
Schr\"{o}dinger equation
\begin{equation}
\Bigl[-\frac{\hbar^2}{2m}\frac{d^2}{dx^2}+V(x)\Bigr]\psi(x)=E\psi(x)
\end{equation}
is the asymptotic expansion of the wave function $\psi(x)$ with
respect to the small Plank constant $\hbar$. Within the standard
semiclassical approach, when the logarithm of the wave function is
expanded
\begin{equation}
\ln \psi(x)=\sum_{n=-1}^{\infty}\hbar^nS_n,
\end{equation}
the treatment of higher orders encounters difficulties because the
recurrence relation, from which the $n$-th term $S_n$ is obtained,
involves in a nonlinear manner all previous terms $S_k$, $k<n$
\cite{LandQM}. To avoid this difficulty, the Milne transformation
should be used, which reduces the problem to the study of
semiclassical expansion of the 'wavelength' $\lambda(x)$, related
to the wave function by
\begin{equation}
\psi(x)=\sqrt{\lambda(x)}\sin\Bigl[\int^x\frac{dx'}{\lambda(x')}\Bigr],
\label{5.1.3}
\end{equation}
and satisfies the third order linear differential equation (see
also Sec.4.1) \cite{Sol84}
\begin{equation}
-\frac{\hbar^2}{4m}\frac{d^3\lambda(x)}{dx^3}+2[V(x)-E]\frac{d\lambda(x)}{dx}+\frac{dV(x)}{dx}\lambda=0.
\label{5.1.4}
\end{equation}
In this approach the coefficients of the expansion
\begin{equation}
\lambda(x)=\sum_{n=0}^{\infty}\hbar^{2n}\lambda_n(x)
\end{equation}
are subject to the two-term linear recurrence relation
\begin{equation}
\lambda_n(x)=\frac{1}{\sqrt{p(x)}}\int^x\frac{dx'}{\sqrt{p(x')}}\frac{d^3\lambda_{n-1}(x')}{dx'}+\frac{C_n}{\sqrt{p(x)}},
\label{5.1.6}
\end{equation}
where $p(x)=\sqrt{2m[E-V(x)]}$ is classical momentum and
$\lambda_0(x)=1/p(x)$. The constant $C_n$ is determined by a
nonlinear Milne equation; the whole effect of constants $C_n$ in
recurrence relation (\ref{5.1.6}), as well as in solution of
equation (\ref{5.3.2}), is that their change results only in the
change of the normalization factor (in detail see \cite{Sol84})
and, without loss of generality, they will further be omitted.

Now let us consider the divergence of the semiclassical expansion
for a particle with zero energy in a power-law potential taken in
the form $V(x)=-\alpha^2x^{2\nu}/2m$, where $\alpha$ and $\nu$ are
arbitrary constants. This is an interesting case in that it
describes all types of violations of the semiclassical
approximation because of power-law singularities. The value
$\nu=1/2$ corresponds to an ordinary turning point. From the
recurrence relation (\ref{5.1.6}) it is easy to find the general
term
\begin{equation}
\lambda_n(x)=\frac{(-1)^n\Gamma\bigl(n+\frac{\nu/2}{\nu+1}\bigr)\Gamma\bigl(n+\frac{1}{2}\bigr)
\Gamma\bigl(n+\frac{\nu+2}{2(\nu+1)}\bigr)} {\alpha
n!\Gamma\bigl(\frac{\nu}{2(\nu+1)}\bigr)\Gamma(\frac{1}{2}\bigr)\Gamma\bigl(\frac{\nu+2}{2(\nu+1)}\bigr)}\Bigl(\frac{\nu+1}{\alpha
x^{\nu+1}}\Bigr)^{2n}x^{-\nu}, \label{5.1.7}
\end{equation}
From this expression follows that for $\nu<-1$ the semiclassical
expansion breaks down as $x \to \infty$, while for $\nu>-1$ it
breaks down as $x\to 0$; correspondingly, it becomes exact in the
limit  $x\to 0$ in the first case and in the limit $x \to \infty$
in the second. When $\nu=-1\pm 1/(2q+1)$ ($q=0,1,2,...$), the
expansion is cut off at $n=q$, and summation of the first $q$
terms gives the exact solution of the Schr\"{o}dinger equation in
the form (\ref{5.1.3}).

Expansion  (\ref{5.1.7}) diverges as $(n!)^2$, so that the
corrections $\lambda_n$ can be taken into account only until they
begin to grow. The index $N$ of the order at which the expansion
should be truncated is determined by setting the derivative of
$\lambda_n$ with respect to $n$ equal to zero. At large $n$ the
following result is obtained $N=\alpha x^{\nu+1}/\hbar(\nu+1)$,
which can be rewritten as
\begin{equation}
N=\frac{1}{\hbar}\int_{x_0}^xp(x')dx', \label{5.1.8}
\end{equation}
where $x_0$ is the point where the semiclassical approximation
breaks down ($x_0=0$ if $\nu>-1$ or $x_0=\infty$ if $\nu<-1$).
Thus, the critical index $N$ is equal (in terms of $\hbar$) to the
classical action from the point $x$ to the singularity $x_0$; it
is invariant in form and is found for arbitrary power-law
singularities.

\subsection{Renormgroup symmetry}

The semiclassical expansion can be written, as well, in the form
\cite{Din,Rak89}
\begin{equation}
\psi^{\pm}(x)=\frac{1}{\sqrt{p(x)}}e^{\pm
iS(x)/\hbar}\sum_{n=0}^{\infty}\Bigl[\pm \frac{i\hbar}{2}\Bigr]^n
\phi_n(x),
\label{5.3.1}
\end{equation}
where $S(x)$ is the classical action calculated from the singular
point $x_0$. Further, it is convenient to take the action $S(x)$
as a new independent variable instead of the coordinate $x$. Then,
the substitution of (\ref{5.3.1}) into the Schr\"{o}dinger
equation leads to the recurrence relation
\begin{equation}
\frac{d\phi_n(S)}{dS}=\frac{d^2\phi_{n-1}(S)}{dS^2}+Q(S)\phi_{n-1}(S),
\label{5.3.2}
\end{equation}
where
\begin{equation}
Q(S)=-\frac{1}{\sqrt{p}}\frac{d^2\sqrt{p}}{dS^2}. \label{5.3.3}
\end{equation}
and the leading term is $\phi_0(x)=1$. Since $Q(x)$  has a pole of
second order at the singular point $x_0$ with residual
$\nu(\nu+2)/[2(\nu+1)]^2$, the solution of the recurrence relation
(\ref{5.3.2}) has a pole of $n$-th order with factorial growing
residual:
\begin{equation}
\phi_n(S)\approx(-1)^n \frac{(n-1)!}{S^n}. \label{5.3.4}
\end{equation}
To analyze higher-order terms, let us introduce asymptotic
expansion of $\phi_n(S)$ in inverse powers of $n$ in the following
special form:
\begin{eqnarray}
\phi_n(S)=(-1)^n\Bigl[\frac{\Gamma(n)}{S^n}\phi^{(1)}_0(S)+
\frac{\Gamma(n-1)}{S^{n-1}}\phi^{(1)}_1(S) ... \nonumber
\\
...+\frac{\Gamma(n-k)}{S^{n-k}}\phi^{(1)}_k(S)+...\Bigr]
\label{5.3.5}
\end{eqnarray}
with the initial condition $\phi_0^{(1)}(x)=1$. Then, from
equation $(\ref{5.3.2})$, for $\phi^{(1)}_k(S)$ the recurrence
relation is obtained
\begin{equation}
\frac{d\phi^{(1)}_k(S)}{dS}=\frac{d^2\phi^{(1)}_{k-1}(S)}{dS^2}+Q(S)\phi^{(1)}_{k-1}(S),
\end{equation}
which is exactly the same as $(\ref{5.3.2})$. Moreover, if one
expands the function $\phi^{(1)}_k(S)$ for large $k$ in the same
manner
\begin{eqnarray}
\phi_k^{(1)}(S)=(-1)^k\Bigl[\frac{\Gamma(k)}{S^k}\phi^{(2)}_0(S)+
\frac{\Gamma(k-1)}{S^{k-1}}\phi^{(2)}_1(S) ... \nonumber
\\
...+\frac{\Gamma(k-s)}{S^{k-s}}\phi^{(2)}_s(S)+...\Bigr],
\label{5.3.7}
\end{eqnarray}
then the new function $\phi^{(2)}_s(S)$ will again be subject to
the recurrence relation (\ref{5.3.2}). We can repeat this
procedure $m$ times and obtain the same recurrence relation, i.e.,
the same sequence $\{\phi^{(m)}_n(S)\}$ as $\{\phi_n(S)\}$. This
amazing property reminds  a self-similarity structure appearing in
the theory of dynamical systems where the picture of irregular
(chaotic) motion repeats itself as one goes to finer scales of
phase space - so called renormgroup symmetry \cite{Ott}. This
property was discovered by Dingle (see \cite{Din}, Chapter XIII)
and independently in \cite{Rak89}.

\subsection{Criterion of accuracy}

With increasing $n$ the terms of the asymptotic expansion
(\ref{5.3.1}) at first decrease until for some number $N$ they
reach the minimal value, after which they increase. The accuracy
of the finite sum is determined by the last retained term.
Therefore, it is correct to sum up only first $N$ terms; the
addition of the higher terms would make the result worse. As in
Sec.5.4.1, the critical index $N$ is determined by setting the
derivative of $\hbar^n \phi_n(x)$ with respect to $n$ equal to
zero. At large values of $n$, using (\ref{5.3.4}) and Stirling
approximation for the $n!$, it follows
\begin{equation}
N=\frac{2S}{\hbar}. \label{5.3.8}
\end{equation}
i.e., the classical action up to nearest (in terms of  $S$)
singularity determines the number of terms of expansion
(\ref{5.3.1}) for best accuracy. Taking into account that the
expansion in Sec.5.4.1 is over even powers of $\hbar$, one can see
that (\ref{5.1.8}) and (\ref{5.3.8}) coincide, but the result
(\ref{5.3.8}) is obtained in the general case. The relative error
is estimated by
\begin{equation}
\Bigl(\frac{\hbar}{2}\Bigr)^N\frac{\phi_N(x)}{\phi_0(x)}\sim\frac{\hbar}{S}e^{-2S/\hbar},
\end{equation}

It is also interesting to see how many terms it is allowed to keep
in asymptotic sub-expansion (\ref{5.3.5}). For large $k$ we can
replace $\phi_k^{(1)}(x)$ by the leading term of expansion
(\ref{5.3.7}). Then, it can be verified in the same manner as for
derivation (\ref{5.3.8}) that the minimal term in sub-expansion
(\ref{5.3.5}) is at $k=n/2$. The same result is obtained for all
next sub-expansions $\phi_s^{(m)}(x)$ ($m=2,3,4,...$).

\section{Concluding remarks; classical physics and measurement}
We obtain any data from measurement which, in principle, gives the
result in terms of classical physics. Intuitively, it is clear
what the concept 'measurement' means; however, it cannot be
formalized, i.e. to be written in mathematical form. What is the
relation between quantum theory and measurement? In classical
mechanics and classical electro-dynamics the concepts 'material
points' and 'electro-magnetic waves' were introduced. In this
sense we can consider quantum theory as a theory of 'information
field' which is described by wave function $\Psi({\bi r })$
\cite{Sol09}; it is additional to 'material points' and
'electro-magnetic waves' substances but without mass and energy.
The problem of the existence of 'information field' is similar to
the situation with electro-magnetic field - nobody suspected its
existence before classical electro-dynamics. {\it Experimentum
crucis}, which proves the existence of information field, is based
on the Einstein-Rozen-Podolsky phenomenon. One of the realizations
of this experiment is the following. After a radiative decay of a
hydrogen atom from a metastable $2S$-state two photons are emitted
having spins with opposite directions
\begin{equation}
{\bi s}_1=-{\bi s}_2, \label{6.1.1}
\end{equation}
since the initial ($2S$) and final ($1S$) states of hydrogen atom
have angular momentum equal to zero. However, the orientation of
each spin is uncertain in the same manner as a position of a
particle in a potential well. If the measurement of the direction
of the spin of the first photon is performed the second photon
takes the opposite direction of spin, according to (\ref{6.1.1}),
independently of the distance between photons. The actual fixation
of the spin orientation of the second photon happens immediately
and it is an actual changing of its state (Bell's inequality
\cite{Bell}). It does not contradict the relativistic restriction
$v\leq c$ because the carrier of information is not material, i.e.
it has no relation to the transfer of mass or energy. Recent
experiment \cite{Exp} demonstrates that the speed of quantum
information is at least $10^{4}$ greater than the speed of light
$c$. This experiment confirms that an information field does
exist. Notice that this interpretation is in contradiction with
wave-particle duality: the wave-type Schr\"odinger equation
describes just 'information field' $\Psi({\bi r})$. Thus, there is
no duality because of two different physical substances -
particles and information field. The interaction between particles
and information fields is measurement which cannot be formalized,
i.e. to be written in mathematical form.

\section*{Acknowledgment}

I am grateful to John Briggs and Tasko Grozdanov for valuable
comments.

\end{document}